\documentclass[11pt]{article}
\usepackage{jheppub}
\usepackage{amsmath}
\usepackage{graphicx}
\usepackage[usenames,dvipsnames,table]{xcolor}
\usepackage{amssymb}
\usepackage{float}
\usepackage{scrtime}
\usepackage{fancyhdr}
\usepackage{subfig}
\usepackage{lipsum}
\usepackage{rotating}
\usepackage{floatpag}
\usepackage{varioref}
\usepackage{slashed}
\usepackage{enumerate}
\usepackage{listings}
\usepackage{cancel}
\usepackage{wrapfig}
\usepackage{mathtools}
\usepackage{pdfpages}
\usepackage[toc,page]{appendix}
\usepackage{multirow}
\usepackage{soul}

\usepackage{caption}
\definecolor{darkgreen}{rgb}{0.0, 0.5, 0.0}

\newcommand{\code}[1]{{\small{\texttt{#1}}}}

\newcommand{\ra}{\ensuremath{\rightarrow}}
\newcommand{\Ra}{\ensuremath{\Rightarrow}}

\usepackage{pifont}% http://ctan.org/pkg/pifont

\def\mg{M_{\tilde{g}}}
\newcommand\Mchipm[1]{M_{\widetilde{\chi}_{#1}^\pm}}

\def\Mchi0{M_{\widetilde{\chi}_i^0}}

\labelformat{section}{Section #1}
\labelformat{subsection}{Section #1}
\labelformat{equation}{Eq.~(#1)}
\labelformat{figure}{Fig.~#1}
\labelformat{subfig}{Fig.~\thefigure#1}
\labelformat{table}{Tab.~#1}
\labelformat{lstlisting}{Listing~#1}

\def\sin{\hspace{.5mm}\text{sin}\hspace{.5mm}}

\def\cos{\hspace{.5mm}\text{cos}\hspace{.5mm}}

\def\tan{\hspace{.5mm}\text{tan}\hspace{.5mm}}

\def\tanb{\hspace{.5mm}\text{tan}\hspace{0.5mm}\beta}
\def\cotb{\hspace{.5mm}\text{cot}\hspace{0.5mm}\beta}

\def\tana{\hspace{.5mm}\text{tan}\hspace{0.5mm}\alpha}
\def\cota{\hspace{.5mm}\text{cot}\hspace{0.5mm}\alpha}

\def\t{\theta}

\def\g{\gamma}

\def\lam{\lambda}
\def\a{\alpha}
\def\b{\beta}
\def\mg{M_{\tilde{g}}}

\def\be{\begin{equation}}
\def\ee{\end{equation}}
\def\bea{\begin{eqnarray}}
\def\eea{\end{eqnarray}}
\def\nn{\nonumber}

 \title{Threshold Corrections to the Bottom Quark Mass Revisited}

 \author[a,b]{Archana Anandakrishnan,}
 \author[a]{B. Charles Bryant,}
 \author[a]{and Stuart Raby}

 \affiliation[a]{\em Department of Physics, The Ohio State University Columbus, OH 43210, USA  \medskip}

 \affiliation[b]{\em Laboratory for Elementary Particle Physics, Cornell University, Ithaca, NY 14853, USA}

 \emailAdd{anandakrishnan.1@osu.edu}
 \emailAdd{bryant.1509@osu.edu}
 \emailAdd{raby.1@osu.edu}

 \abstract{ %\normalsize\parindent 0pt\parskip 0pt
 Threshold corrections to the bottom quark mass are often estimated under the approximation that $\tanb$ enhanced contributions are the most dominant. In this work we revisit this common approximation made to the estimation of the supersymmetric threshold corrections to the bottom quark mass. We calculate the full one-loop supersymmetric corrections to the bottom quark mass and survey a large part of the phenomenological MSSM parameter space to study the validity of considering only the $\tanb$ enhanced corrections. Our analysis demonstrates that this approximation underestimates the size of the threshold corrections by $\sim$12.5\% for most of the considered parameter space. We discuss the consequences for fitting the bottom quark mass and for the effective couplings to Higgses. We find that it is important to consider the additional contributions when fitting the bottom quark mass but the modifications to the effective Higgs couplings are typically $\mathcal{O}$(few)\% for the majority of the parameter space considered.
 }

\begin{document}
 \maketitle

	\section{Introduction \label{Sec:intro}}
	
	The supersymmetric (SUSY) threshold corrections to the bottom quark mass in the large $\tanb$ regime are often expressed as an approximation of the dominant gluino-sbottom and chargino-stop loop contributions~\cite{Hall:1993gn, Carena:1994bv, Blazek:1995nv},
	
	\bea
	\left(\frac{\Delta m_b}{m_b}\right)^{\rm app} = 
	&&\frac{8}{3} \frac{g_3^2}{16 \pi^2}\mg (\mu\tanb-A_b) I(\mg^2, m_{\tilde{b}_1}^2, m_{\tilde{b}_2}^2)  + \nn \\ 
	&&\frac{\lam^2_t}{16\pi^2}\mu (A_t\tanb-\mu) I(\mu^2,m_{\tilde{t}_1}^2,m_{\tilde{t}_2}^2)
	\label{Eq:common-app}
	\;.\eea
	
	In order to fit the bottom quark mass, $m_b(M_Z)^{\rm SM} = m_b(M_Z)^{\rm MSSM}\left(1 + \Delta m_b/m_b\right)$, where $m_b(M_Z)^{\rm MSSM}$ is obtained from the evolution of the bottom Yukawa coupling from a UV scale (such as the GUT scale) to the $M_Z$ scale. The effects of these supersymmetric threshold corrections are important especially in the era of precision Higgs couplings and flavor physics and has been a part of many analyses. For some recent work, see~\cite{Carena:2014nza, Allanach:2014nba, Altunkaynak:2013xya, Harlander:2012pb, Altmannshofer:2012ks, Elor:2012ig, Barger:2012ky, Joshipura:2012sr, BhupalDev:2012nm}.
	
	Let us first summarize some of the well-known consequences of the above expression for a common type of model that has large $\tanb$ such as models with third family Yukawa unification. In such models, the threshold corrections typically need to be $\mathcal{O}(\text{few})\%$\ and negative. These corrections can often be large thus the two terms in~\ref{Eq:common-app} must either nearly cancel or both be suppressed.
	
	For $\mu>0$ and $\tanb \simeq 50$, $A_t$ must be large and negative in order for the two contributions to approximately cancel and yield a negative value. This in turn has consequences for flavor physics. The branching ratio for $B_s \ra \mu^+ \mu^-$ receives large $\tanb$-enhanced contributions from Higgs-mediated neutral currents that are proportional to $A_t^2(\tanb)^6/M^4_A$~\cite{Choudhury:1998ze, Babu:1999hn}. In order to be in agreement with the experimental value which is measured at $3.2 \times 10^{-9}$, $M_A$ must be large if $A_t$ and $\tanb$ are large. An important constraint to then consider is the inclusive decay $B_s \ra X_s \g$ to which the dominant SUSY contributions are a chargino-stop loop and a top-charged Higgs loop~\cite{Bobeth:1999ww, Degrassi:2000qf, Carena:2000uj}. The chargino contribution is $\tanb$-enhanced and, with large and negative $A_t$, adds destructively to the SM branching ratio. The charged Higgs contribution, on the other hand, adds constructively to the SM branching ratio, but is suppressed by the heavy Higgs masses required to be consistent with $\mathcal{B} (B_s \ra \mu^+ \mu^-)$. Since the SM prediction is in good agreement with the data, these two contributions must nearly cancel. Such a cancellation is difficult to obtain in the given region of parameter space and one is then led to consider heavy scalars~\cite{Albrecht:2007ii}.
	
	The situation is different for $\mu<0$ since the gluino contribution, which is the dominant contribution, already has the needed sign. In this case, the parameters need not be large in order to obtain a small threshold correction. This region of parameter space however was initially disfavored due to conflicts with flavor physics. When $\mu < 0$, the chargino contributions add constructively with the SM contributions to the $\mathcal{B}(B_s\ra X_s\g)$ observable and hence yield enhanced values~\cite{Baer:1997jq,Baer:2000jj,Carena:2000uj,Auto:2003ys}. Additional complications also arise due to tensions with the $(g-2)_\mu$ observable in this regime, where the theoretical prediction is too small to match the experimental value. More recently, viable models with $\mu<0$ have been constructed but they typically have squark masses greater than 1 TeV~\cite{Badziak:2013eda, Ajaib:2014ana, Anandakrishnan:2013cwa}.
	
	Fitting the bottom quark mass and satisfying current experimental constraints from flavor physics has therefore pushed Yukawa unified models into the territory of heavy scalars. Other models may of course be constructed that evade such restrictions, but the absence of the detection of any new physics at the LHC generically requires one to consider heavy scalar masses. The current limits on the colored superpartner masses are already approaching the TeV range~\cite{ATLAS-SUSY, CMS-SUSY}. As we transition into the TeV region of the SUSY parameter space, a re-evaluation of the approximations of SUSY threshold corrections to the bottom quark is warranted. This is especially important in the era of precision physics since the approximation is often invoked in studies of bottom quark mass and couplings.
	
	In order to understand the size and behavior of the threshold corrections to the bottom quark, we survey a large part of the parameter space of interest and choose to scan over the parameters of the pMSSM instead of restricting ourselves to a particular model. For each point, we calculate both the full, exact one-loop radiative corrections to the bottom quark and compare with the value obtained from the approximate form of the corrections as given in~\ref{Eq:common-app}. For each point in the pMSSM scan, we additionally check the Higgs mass and constraints from $\mathcal{B} (B_s \ra X_s \gamma)$ and $\mathcal{B}(B \ra \mu^+ \mu^-)$.
	
	This paper is organized as follows. The details of the parameter scan are presented in~\ref{Sec:scan}. In~\ref{Sec:ex-app}, we present the full, exact one-loop corrections compared to the approximate form of the contributions and motivate the need for a scrutiny of this approximation. We then consider in turn each approximation made to the individual contributions to the threshold correction in~\ref{Sec:indy}.~\ref{Sec:consequences} surveys the consequences of using the full expression of the threshold corrections to the bottom quark. Finally, we conclude in~\ref{Sec:conclusions}.
		
% % % % % % % % % % % % % % % % % % % % % % % % % % % % % % % % % % % % % % % % %
% % % % % % % % % % % % % % % % % % % % % % % % % % % % % % % % % % % % % % % % %	

%	\newpage
	\section{Parameter Scan \label{Sec:scan}}
	
    The pMSSM parameter space is defined by $\{m_{Q_i}, m_{u_i}, m_{d_i}, m_{L_i}, m_{e_i}, A_i, M_i, M_A, \mu, \tanb\}$ with the family index $i=1$-$3$. We consider the inter-generational mixing to be negligible and that the masses of the first two family scalars are large relative to the third family scalar masses. We therefore ignore contributions to the bottom quark mass from the first two families. In this analysis, we fix $\tanb=50$ in which region the SUSY threshold corrections are dominant.\footnote{With $\tanb =50$, third family Yukawa unification can also be satisfied.} The ranges for the remaining SUSY parameters are given in~\ref{tab:params}. With these parameter bounds, we randomly generate 50,000 points. We then use \code{micrOMEGAs}~\cite{Belanger:2014hqa} to calculate the quantities $m_h$, $\mathcal{B}(B_s\ra\mu^+\mu^-)$, and $\mathcal{B}(B_s \ra X_s\gamma)$. Only points for which these quantities satisfy current experimental bounds are retained.

    For the Standard Model parameters, we use the measured values of the top quark, W, Z, and Higgs masses. Note that we use $m_h=125.3$ GeV for all points when calculating threshold corrections. After running through \code{micrOMEGAs}, the points that survive all have a Higgs mass within 3 GeV of this value. This is at most a $\sim$2\% difference. Furthermore, the Higgs mass only occurs in the calculation of the neutral Higgs contribution. The error in this approximation is therefore negligible and the results remain unaffected. For the bottom quark mass, we use the \code{RunDec} package~\cite{Chetyrkin:2000yt} to run $m_b(m_b)$ to $m_b(M_Z)$.
    \clearpage
	\begin{table}[h!]
    \centering
    \begin{tabular}{c c c}\hline
    $g_1$ = 0.46 & $g_2$ = 0.64& $g_3$ = 1.2 \\
	$M_t$ = 173.36& $m_b$ = 2.69& $V_{tb}$ = 1 \\
	$M_Z$ = 91.1876& $M_W$ = 80.385& $m_h$ = 125.3 \\
	$v$ = 246& $\tanb$ = 50 &  \\ \hline
	&$1000 < \{m_{Q_3},m_{u_3},m_{d_3}\} < 5000$  &\\
	&$~100 < \{m_{L_3},m_{e_3}\} < 5000$ & \\
	&$-15000 < \{A_t,A_b\} < 15000~~~$ & \\
	&$-1000 < \{M_1,M_2\} < 1000~~$ & \\
	&$~500 < M_3 < 2000$&  \\
	&$~1000 < M_A < 2000$& \\	
	&$-2000 < \mu < 2000~~$& \\ \hline
	\end{tabular}
	\caption{Parameter values and ranges at $M_Z$.\\ All masses in GeV.} \label{tab:params}	
	\end{table}

% % % % % % % % % % % % % % % % % % % % % % % % % % % % % % % % % % % % % % % % %
% % % % % % % % % % % % % % % % % % % % % % % % % % % % % % % % % % % % % % % % %	
	
%	\newpage
	\section{Exact vs. Approximation \label{Sec:ex-app}}
	
	The complete set of one loop corrections to the bottom quark mass is given by~\cite{Pierce:1996zz}
	\be
	\Delta m_b (M_Z) = \Delta m_b^{\tilde{g}} + \Delta m_b^{\tilde{\chi}^\pm} +
	\Delta m_b^{\tilde{\chi}^0} + \Delta m_b^{H^\pm} + \Delta m_b^{A} +
	\Delta m_b^{h}  + \Delta m_b^{W} + \Delta m_b^{Z}
	\;,\ee
	with the tree level mass given by $\lambda_b (M_Z) \frac{v}{\sqrt{2}} {\rm cos} \beta$.	

	\begin{figure}[h!]
	\thisfloatpagestyle{empty}
	\centering
	\includegraphics[width=0.6\textwidth,  clip]{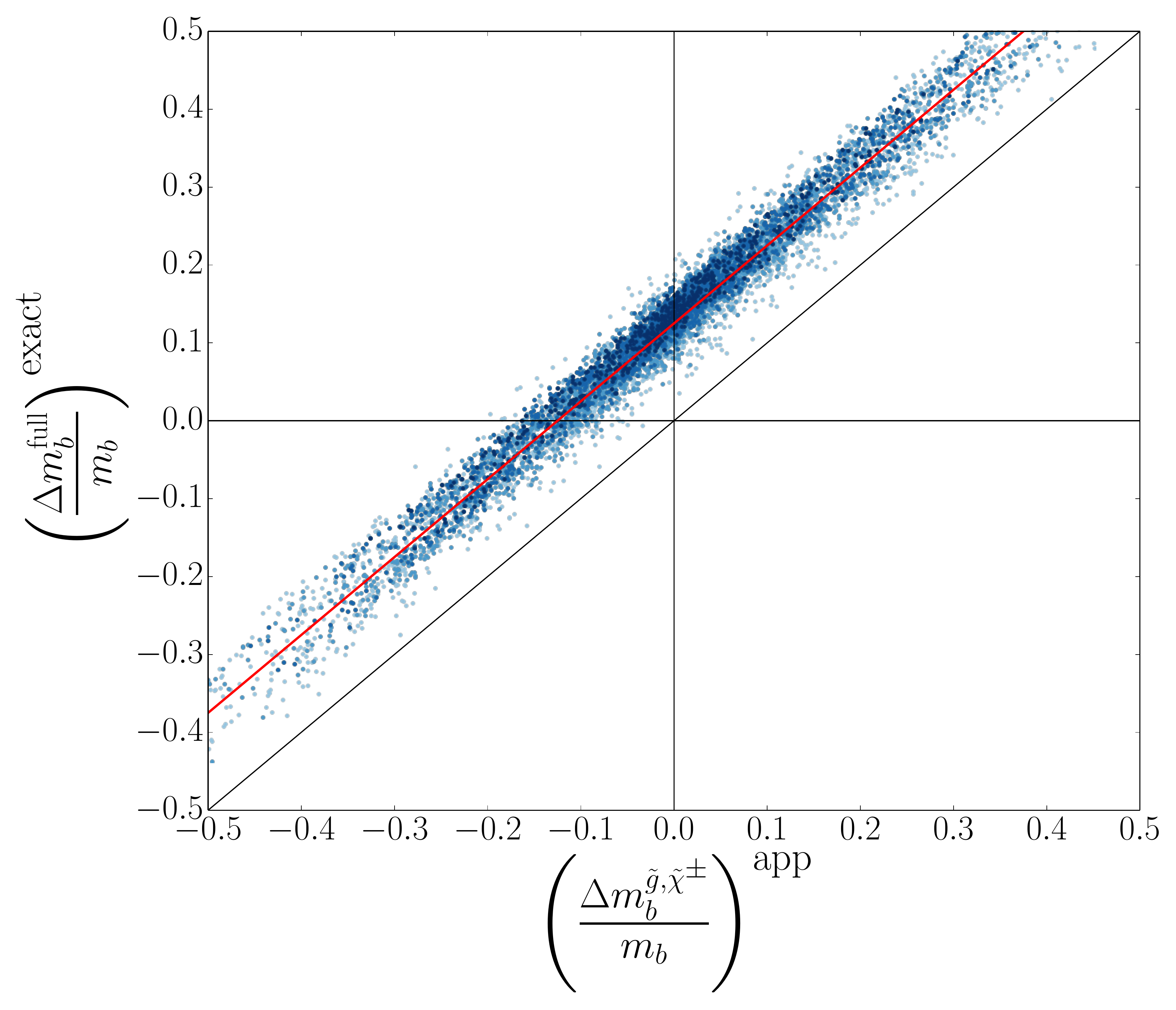}
	\caption{\small The plot shows the full, exact one-loop threshold corrections to the bottom quark mass vs. the approximate form of the correction given in~\ref{Eq:common-app}. Darker shades of blue represent increasing squark masses from 1 TeV to $\ge$4 TeV. The black (lower) diagonal line represents where the exact and approximate forms would be equal. The red (upper) diagonal line represents where the correction from the exact form is $\sim$12.5\% larger than the correction from the approximate form. 
	}
	\label{fig:ex-app}
	\end{figure}

	In~\ref{fig:ex-app}, we present the results of the parameter scan by plotting the full, exact one-loop threshold corrections to the bottom quark mass against the approximate form of the corrections given in~\ref{Eq:common-app}. The color gradient represents squark masses from 1 TeV at the lightest to $\ge$4 TeV at the darkest. The black (lower) diagonal line represents where the exact and approximate forms would be equal. The red (upper) diagonal line is to help guide the eye and represents where the correction from the exact form is $\sim$12.5\% larger than the correction from the approximate form. All of the points lie along the latter line and thus there is a nonnegligible difference between the exact and approximate forms of the threshold correction in this region of parameter space. We now consider the individual contributions in turn to discover the source(s) of the discrepancy.
	
% % % % % % % % % % % % % % % % % % % % % % % % % % % % % % % % % % % % % % % % %
% % % % % % % % % % % % % % % % % % % % % % % % % % % % % % % % % % % % % % % % %
		
%	\newpage
	\section{Individual contributions \label{Sec:indy}}
	
	\subsection{Gluino-Sbottom \label{Sec:gl-sbot}}
	
	We look first at the approximation made to the gluino-sbottom contribution. Gluinos couple to the down-type squarks and quarks proportional to the SU(3) gauge coupling $g_3$ and hence contribute large corrections to the bottom quark mass. The corrections are dominant when the squarks belong to the third family since the inter-generational mixings between the squarks are typically (and by assumption in this study) small. The detailed calculation can be found in the appendix. We quote the final, exact form here~\cite{Pierce:1996zz}.
	
	\bea
	\Delta m_b^{\tilde{g}} = \frac{8}{3} \frac{g_3^2}{16 \pi^2} \left[ \frac{\sin 2\theta_b \mg }{2} \left(B_0 \left(p, \mg, m_{\tilde{b}_1}\right)
	- B_0 \left(p, \mg, m_{\tilde{b}_2}\right)  \right)
	\right. \nn \\ \left.
	- \frac{m_b}{2} \left(B_1 \left(p, \mg, m_{\tilde{b}_1}\right) +
	B_1 \left(p, \mg, m_{\tilde{b}_2}\right) \right) \right]
	\label{Eq:fullgluino}
	\;,\eea

	where the momentum of the bottom quark is given by $p$. In the limit $p \rightarrow 0$ (which is a good assumption here since $p^2 = m_b^2$), the Passarino-Veltman functions can be written as
	\bea
	B_0(0, \mg, m_{\tilde{b}}) &=& - \ln \left(\frac{m_{\tilde{b}}^2}{Q^2}\right) +
	1 + \left(\frac{1}{1 - x}\right) \ln  x \\
	B_1(0, \mg, m_{\tilde{b}}) &=& \frac{1}{2} \left[-\ln \left(\frac{m_{\tilde{b}}^2}{Q^2}\right) + \frac{1}{2} + \frac{1}{1-x} + \frac{\ln x}{(1-x)^2} -\t(1-x)\ln x\right] \label{Eq:pass-velt-B1}
	\eea
	where $x = m_{\tilde{b}}^2/\mg^2$. The first term in the above expression simplifies to
	\bea
	&&\frac{\sin 2\theta_b \mg }{2} \left[B_0 \left(p, \mg, m_{\tilde{b}_1}\right) - B_0 \left(p, \mg, m_{\tilde{b}_2}\right)  \right] \nonumber \\
	&=& \frac{\sin 2\theta_b \mg }{2} \left[ \ln \left(\frac{m_{\tilde{b}_2}^2}{m_{\tilde{b}_1}^2}\right) + \mg^2 \left(\frac{1}{\mg^2 - m_{\tilde{b}_1}^2}
	\ln  \left(\frac{m_{\tilde{b}_1}^2}{\mg^2}\right) - \frac{1}{\mg^2 - m_{\tilde{b}_2}^2}
	\ln  \left(\frac{m_{\tilde{b}_2}^2}{\mg^2}\right) \right) \right]
	\;.
	\eea
    	
	The angle $\sin 2\theta_b$ can be determined to be
	\be
	\sin 2\theta_b = \frac{2 m_b (\mu\tanb -A_b)}{\sqrt{(m_{\tilde{b}_L}^2 - m_{\tilde{b}_R}^2)^2 + (2 m_b (\mu\tanb -A_b))^2}}
	= \frac{2m_b (\mu\tanb -A_b)}{m_{\tilde{b}_2}^2 - m_{\tilde{b}_1}^2}
	\;,\ee
	where we have ignored terms proportional to $M_Z$ or $m_b$. The trilinear coupling $A_b$ is often ignored since $\mu$ is enhanced by $\tanb$.\footnote{We keep $A_b$ here in order to be consistent with the definitions of the squark masses.} Similarly, the second term in~\ref{fullgluino} is also neglected. Collecting terms, we arrive at the form in~\ref{Eq:common-app},
	\be
	\frac{\Delta m_b^{\tilde{g}}}{m_b} \simeq
	\frac{8}{3} \frac{g_3^2}{16 \pi^2}\mg (\mu\tanb-A_b) I(\mg^2, m_{\tilde{b}_1}^2, m_{\tilde{b}_2}^2)
	\;,
	\label{Eq:gluino-app}
	\ee
	where
	\be
	I (a, b, c) = \frac{a b \ln \left(\frac{a}{b}\right) + b c \ln \left(\frac{b}{c}\right) + a c \ln \left(\frac{c}{a}\right)}{(a-b)(b-c)(a-c)}
	\;.
	\ee
	
	This is the expression that is typically used in most of the literature with large $\tanb$ models. 
	
	\begin{figure}[h!]
	\thisfloatpagestyle{empty}
	\centering
	\includegraphics[width=0.6\textwidth,  clip]{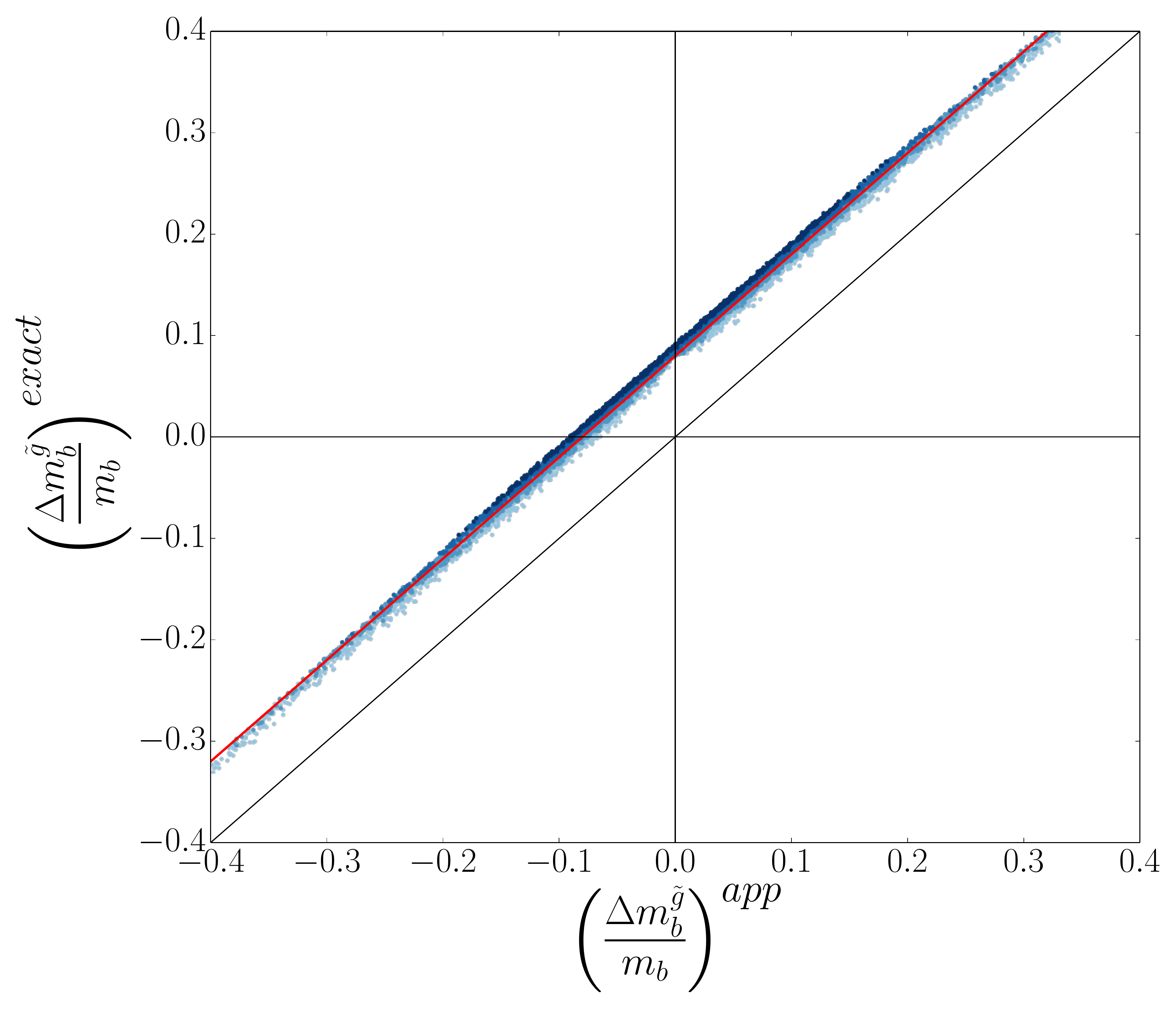}
	\caption{\small The plot shows the exact, one-loop gluino-sbottom threshold correction to the bottom quark mass vs. the approximate form of this correction given in~\ref{Eq:common-app}. Darker shades of blue represent increasing squark masses from 1 TeV to $\ge$4 TeV. The black (lower) diagonal line represents where the exact and approximate forms would be equal. The red (upper) diagonal line represents where the correction from the exact form is $\sim$8\% larger than the correction from the approximate form. %As the squark masses increase, the approximations begin to worsen.
	}
	\label{fig:gl-ex-app}
	\end{figure}	
	
	In~\ref{fig:gl-ex-app}, the exact, one-loop gluino-sbottom threshold correction to the bottom quark mass is compared to the approximate form of this correction given in~\ref{Eq:common-app}. Darker shades of blue represent increasing squark masses from 1 TeV to $\ge$4 TeV. The black (lower) diagonal line represents where the exact and approximate forms would be equal. The red (upper) diagonal line represents where the correction from the exact form is $\sim$8\% larger than the correction from the approximate form. Because the approximate form of the gluino-sbottom correction is equal to the terms in the exact form proportional to the $B_0$ Passarino-Veltman functions the discrepancy must be due to the terms in the exact form proportional to the $B_1$ Passarino-Veltman functions. 

	\begin{figure}[h!]
	\thisfloatpagestyle{empty}
	\centering
	\includegraphics[width=0.56\textwidth, clip]{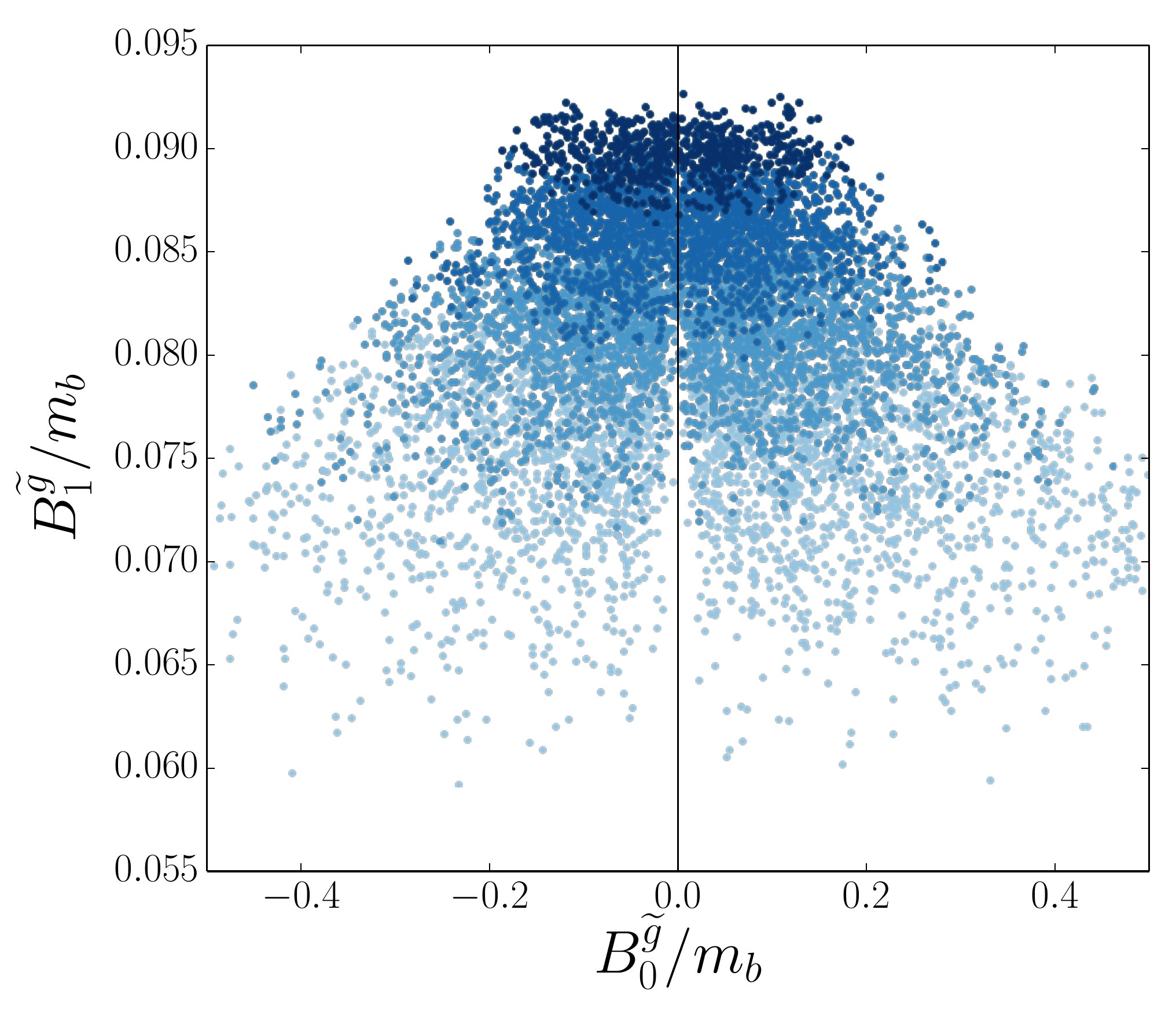}
	\caption{\small We plot the $B_0^{\widetilde{g}}$ term against the $B_1^{\widetilde{g}}$ term ($B_0^{\widetilde{g}}$ and $B_1^{\widetilde{g}}$ are defined in the text). Darker shades represent increasing sbottom masses. As the sbottom masses increase from 1 TeV to $\ge$4 TeV, the $B_0^{\widetilde{g}}$ terms becomes smaller and the two terms are nearly the same magnitude. Furthermore, the points along the vertical line have $(\mu\tanb -A_b)\simeq0$.}
	\label{fig:gl-B0B1}
	\end{figure}	

	We refer to the term in~\ref{Eq:fullgluino} containing the $B_{0(1)}$ Passarino-Veltman functions and its prefactor as the $``B_{0(1)}^{\widetilde{g}}"$ term. In~\ref{fig:gl-B0B1}, the $B_0$ term is plotted against the $B_1^{\widetilde{g}}$ term. The color gradient from light to dark represents increasing sbottom masses from 1 TeV to $\ge$4 TeV. As the sbottom masses get pushed toward more than a few TeV, the $B_0^{\widetilde{g}}$ term decreases while the $B_1^{\widetilde{g}}$ term slightly increases, and the two terms are nearly the same magnitude. The increase in the $B_1^{\widetilde{g}}$ term can be understood by considering~\ref{Eq:pass-velt-B1} in the limit of large sbottom masses. For a fixed gluino mass\footnote{In this analysis we consider gluinos to have mass $\le$2 TeV.} and in the limit of large $x$, one finds for the $B_1^{\widetilde{g}}$ term,
	\be
	\frac{B_1^{\widetilde{g}}}{m_b} \simeq 	\frac{4}{3} \frac{g_3^2}{16 \pi^2}    \left[\ln\left(\frac{m_{\tilde{b}_1}m_{\tilde{b}_2}}{Q^2}\right) - \frac{1}{2} \right] 
	\;.
	\ee
	Thus the $B_1^{\widetilde{g}}$ term grows logarithmically with increasing sbottom masses, which explains why there appears to be a constant vertical shift of $\sim$8\% from the diagonal line along which the approximation is equal to the exact expression in~\ref{fig:gl-ex-app}. In this regime, where the $B_0^{\widetilde{g}}$ term is small, it is therefore important that the $B_1^{\widetilde{g}}$ term not be ignored. Finally, the points along the vertical line in~\ref{fig:gl-B0B1}  have $(\mu\tanb -A_b)\simeq0$, and so one must be careful to check the size of $A_b$ relative to $\mu\tanb$ also.

% % % % % % % % % % % % % % % % % % % % % % % % % % % % % % % % % % % % % % % % % % 
% % % % % % % % % % % % % % % % % % % % % % % % % % % % % % % % % % % % % % % % % %

%	\newpage
	\subsection{Chargino-Stop \label{Sec:ch-stop}}
		
	We turn now to the approximation made to the chargino-stop contribution. The charginos couple to the up-type squarks and down-type quarks proportional to the SU(2) coupling $g_2$ and the Yukawa couplings $\lam_{t,b}$ with strength depending upon their respective wino-higgsino composition. The corrections dominate when the squarks are from the third family due to CKM suppression of the contributions from the first two families of squarks. The calculation is presented in detail in the appendix. The exact closed form cannot be put into a simplified form as was the case for the gluino-sbottom contribution. This is due to the non-trivial convolution of the elements of the stop mixing matrix, the elements of the chargino mixing matrices, and the weak and Yukawa coupling constants obtained by summing over the left and right stops and the two charginos. We therefore list the exact results from the appendix and discuss the approximations made to obtain the form in~\ref{Eq:common-app}.
	
    The full expression is~\cite{Pierce:1996zz}
    \bea
    \Delta m_b^{\tilde{\chi}^\pm_i} &=& \sum_{i=1}^2 \sum_{x=1}^2 B_{LR_i}^x + \frac{m_{b0}}{2} (A_{L_i}^x + A_{R_i}^x)
    \label{Eq:fullchargino-1st}
    \eea	
	with
    \bea
    B_{LR_i}^x &=& -\frac{\bar{\Phi}^x_i \Phi^x_i \Mchipm{i}}{16 \pi^2} B_0(p,\Mchipm{i},m_{\tilde{t}_x}) \nn \\
    A_{L_i}^x &=&  -\frac{ \left(\Phi^x_i\right)^\dagger \Phi^x_i}{16 \pi^2} B_1(p, \Mchipm{i}, m_{\tilde{t}_x}) \nn \\
    A_{R_i}^x &=&  -\frac{ \left(\bar{\Phi}^x_i\right)^\dagger \bar{\Phi}^x_i}{16 \pi^2} B_1(p,\Mchipm{i},m_{\tilde{t}_x})
    \;.\eea
    Here $i=1,2$ is the chargino index and $x=1,2$ is the stop index.

    The couplings are given by
    \bea
    \Phi^x_i &=& \frac{\lambda_t}{\sqrt{2}} V^\dagger_{i2}
    \left(\Gamma^x_R\right)^\dagger - g_2 V^\dagger_{i1} \left(\Gamma^x_L\right)^\dagger \nn \\
    \bar{\Phi}^x_i &=& \frac{\lambda_b}{\sqrt{2}} U^\dagger_{i2} \Gamma^x_L
    \;,\eea
    where $U,V$ are the chargino mixing matrices and $\Gamma_{L,R}$ are the columns of the stop mixing matrix. The momentum of the bottom quark is given by $p$.

    The terms containing the $B_1$ functions are often neglected and so we focus on the $B_{LR_i}^x$ contributions. Setting $p=0$ and expanding these terms,
    \bea
    B_{LR_i}^x &=& -\frac{\bar{\Phi}^x_i \Phi^x_i \Mchipm{i}}{16 \pi^2} B_0(0,\Mchipm{i},m_{\tilde{t}_x}) \nonumber \\
    &=&\frac{-\Mchipm{i}}{16\pi^2}
    \left[\frac{\lam_b}{\sqrt{2}}U^\dag_{i2} \Gamma^x_L\right]
    \left[\frac{\lam_t}{\sqrt{2}} V^\dag_{i2}
    \left(\Gamma^x_R\right)^\dag - g_2 V^\dag_{i1} \left(\Gamma^x_L\right)^\dag \right] B_0(0,\Mchipm{i},m_{\tilde{t}_x})
    \;.\eea

    Neglecting terms proportional to $g_2$ and summing over the stops and charginos yields
    \bea
    \sum_{i=1}^2 \sum_{x=1}^2 B_{LR_i}^x \simeq&&
    \frac{-\Mchipm{1}}{16\pi^2}
    \left[\lam_b \lam_t U^\dag_{12}V^\dag_{12} \frac{\sin 2\theta_t}{2}\right]
    \left[B_0(0,\Mchipm{1},m_{\tilde{t}_1})-B_0(0,\Mchipm{1},m_{\tilde{t}_2})\right]
    \nn \\
    &+&
    \frac{-\Mchipm{2}}{16\pi^2}
    \left[\lam_b \lam_t U^\dag_{22}V^\dag_{22} \frac{\sin 2\theta_t}{2}\right]
    \left[B_0(0,\Mchipm{2},m_{\tilde{t}_1})-B_0(0,\Mchipm{2},m_{\tilde{t}_2})\right]
    \;.\label{Eq:chargino-B0}
    \eea

\begin{figure}[h!]
	\thisfloatpagestyle{empty}
	\centering
	\includegraphics[width=0.6\textwidth,  clip]{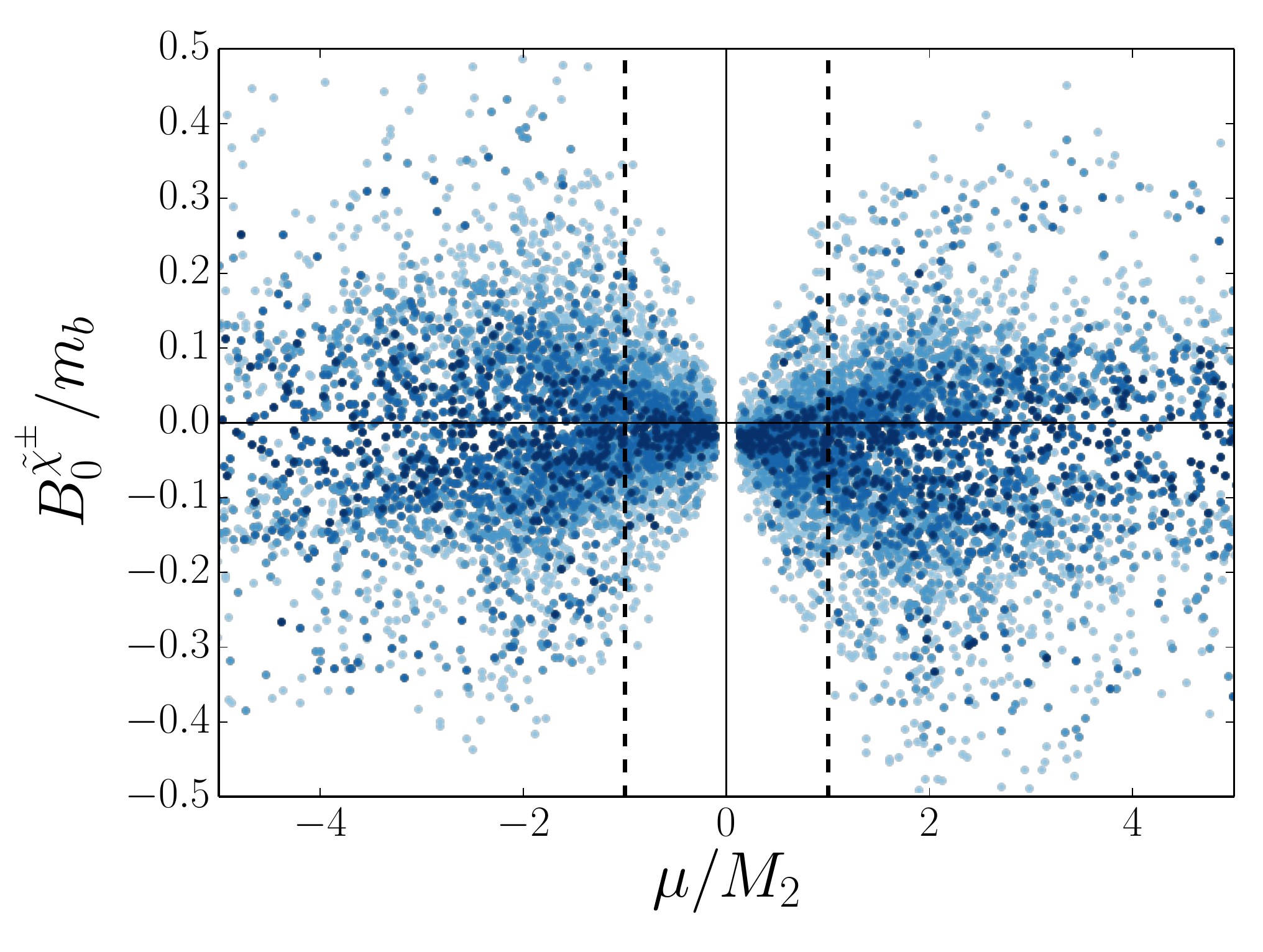}
	\caption{\small The plot shows that the dominant piece $B_0^{\widetilde{\chi}^\pm}$ (defined in the text) of the chargino corrections is small when $|\mu/M_2| < 1$ and can be large when $|\mu/M_2| >> 1$. The vertical dashed lines mark the crossover between these two regimes. Darker shades of blue represent increasing squark masses from 1 TeV to $\ge$4 TeV.}
	\label{fig:ch-muM2}
	\end{figure}

    For $|\mu| > |M_2|$, one finds that $U^\dag_{12}V^\dag_{12}\simeq0$ and $ U^\dag_{22}V^\dag_{22}\simeq1$, whereas for $|\mu| < |M_2|$, one finds that $U^\dag_{12}V^\dag_{12}\simeq1$ and $U^\dag_{22}V^\dag_{22}\simeq0$. Furthermore, $\sin 2\theta_t = -2 \lam_t v_d \tanb(A_t - \frac{\mu}{\tanb})/(m^2_{\tilde{t}_2}-m^2_{\tilde{t}_1})$
    % \simeq 2 \lam_t v_d A_t\tanb/(m^2_{\tilde{t}_2}-m^2_{\tilde{t}_1})$
    so that\footnote{The $\mu/\tanb$ term is often neglected. We keep it here however in order to be consistent with the definitions of the squark masses.}
    \be
    \frac{\Delta m_b^{\tilde{\chi}^\pm_i}}{m_b} \simeq
    \frac{\lam^2_t}{16\pi^2}
    \mu (A_t\tanb -\mu)
    I(\mu^2,m_{\tilde{t}_1}^2,m_{\tilde{t}_2}^2)  
    \;.\label{Eq:chargino-app}
    \ee
    Among the two charginos, the dominant corrections are only from the Higgsino and are proportional to the Higgsino mass, $\mu$. Hence the chargino corrections tend be larger when $|\mu| > |M_2|$ (heavier Higgsino) and smaller when $|\mu| < |M_2|$ (lighter Higgsino) as shown in~\ref{fig:ch-muM2}. We refer to the term in~\ref{Eq:fullchargino-1st} containing the $B_{0(1)}$ Passarino-Veltman functions and its prefactor as the $``B_{0(1)}^{\widetilde{\chi}^\pm}"$ term.

    \begin{figure}[h!]
	\thisfloatpagestyle{empty}
	\centering
	\includegraphics[width=0.6\textwidth,  clip]{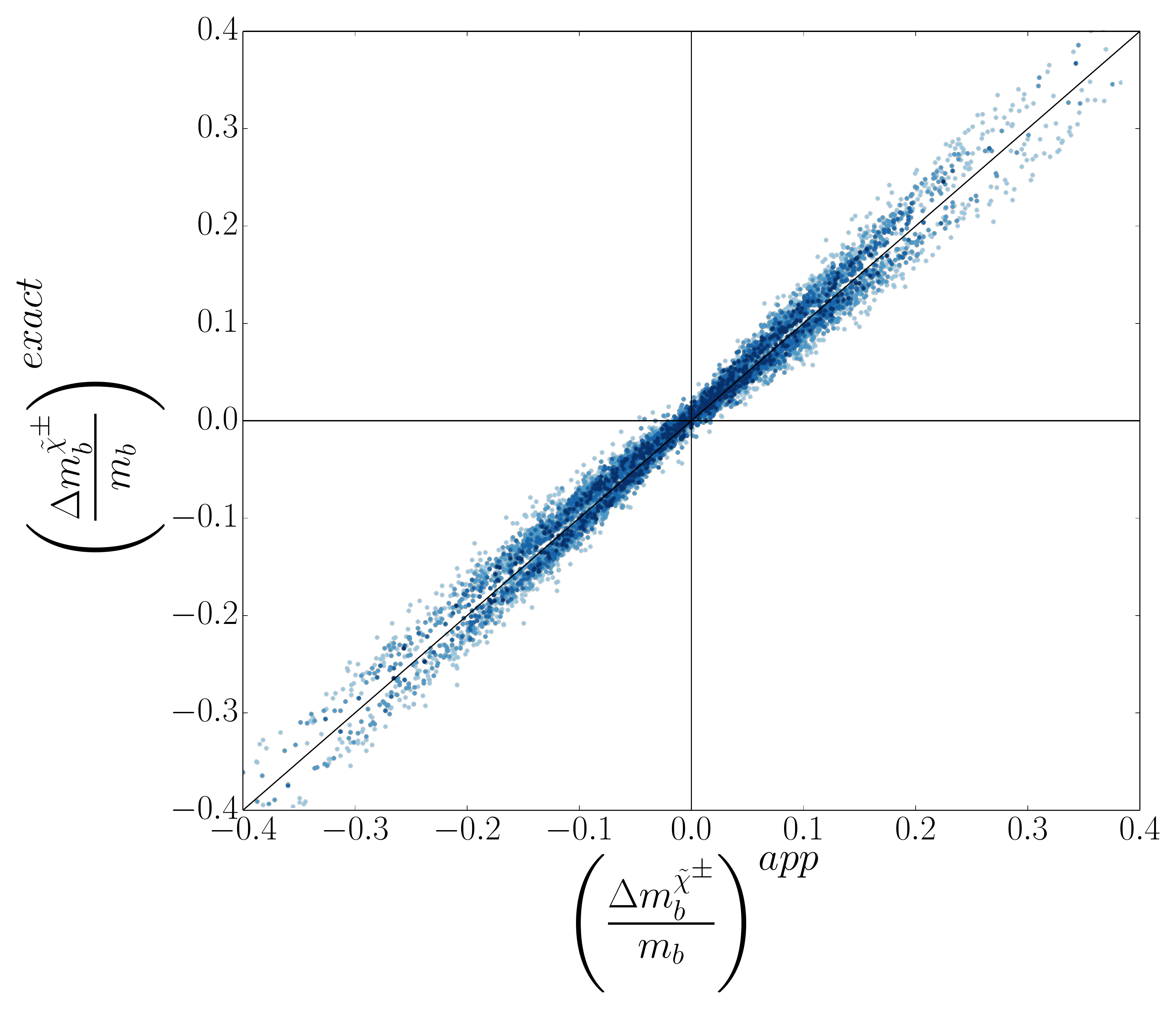}
	\caption{\small The plot shows the exact, one-loop chargino-stop threshold correction to the bottom quark mass vs. the approximate form of this correction given in~\ref{Eq:common-app}. Darker shades of blue represent increasing squark masses from 1 TeV to $\ge$4 TeV. }
	\label{fig:ch-ex-app}
	\end{figure}	
	
	In~\ref{fig:ch-ex-app}, the exact, one-loop chargino-stop threshold correction to the bottom quark mass is compared to the approximate form of this correction given in~\ref{Eq:common-app}. Darker shades of blue represent increasing squark masses from 1 TeV to $\ge$4 TeV. It is clear that the chargino-stop approximation is a good approximation over all of the parameter space, particularly in the region in which the stops are heavy. We note that a nearly constant, positive contribution from the $B_1^{\widetilde{\chi}^\pm}$ term is present as in the gluino-sbottom case. Here however the contribution is $\lesssim2$\% and leaves the chargino-stop approximation as a good approximation.

% % % % % % % % % % % % % % % % % % % % % % % % % % % % % % % % % % % % % % % % %
% % % % % % % % % % % % % % % % % % % % % % % % % % % % % % % % % % % % % % % % %
	
	%\newpage
	\subsection{$W$, $Z$, Higgses, and Neutralinos \label{Sec:rest}}
	
	Due to weaker coupling strengths compared to $g_3$ and $\lam_t$, the contributions to the threshold correction of the bottom quark mass from $W$, $Z$, Higgses, and neutralinos are often neglected. It is possible that while the gluino and chargino contributions may each be of much greater magnitude than these other contributions, a cancellation occurs such that their sum is of the same magnitude as the other contributions. Since these terms are dropped altogether, the validity of this approximation is simply based on the magnitude of their contribution compared to the total approximate correction as given in~\ref{Eq:common-app}.
	
\begin{figure}[h!]
	\centering
	\subfloat[Higgses]{{\includegraphics[width=0.46\textwidth,  clip]{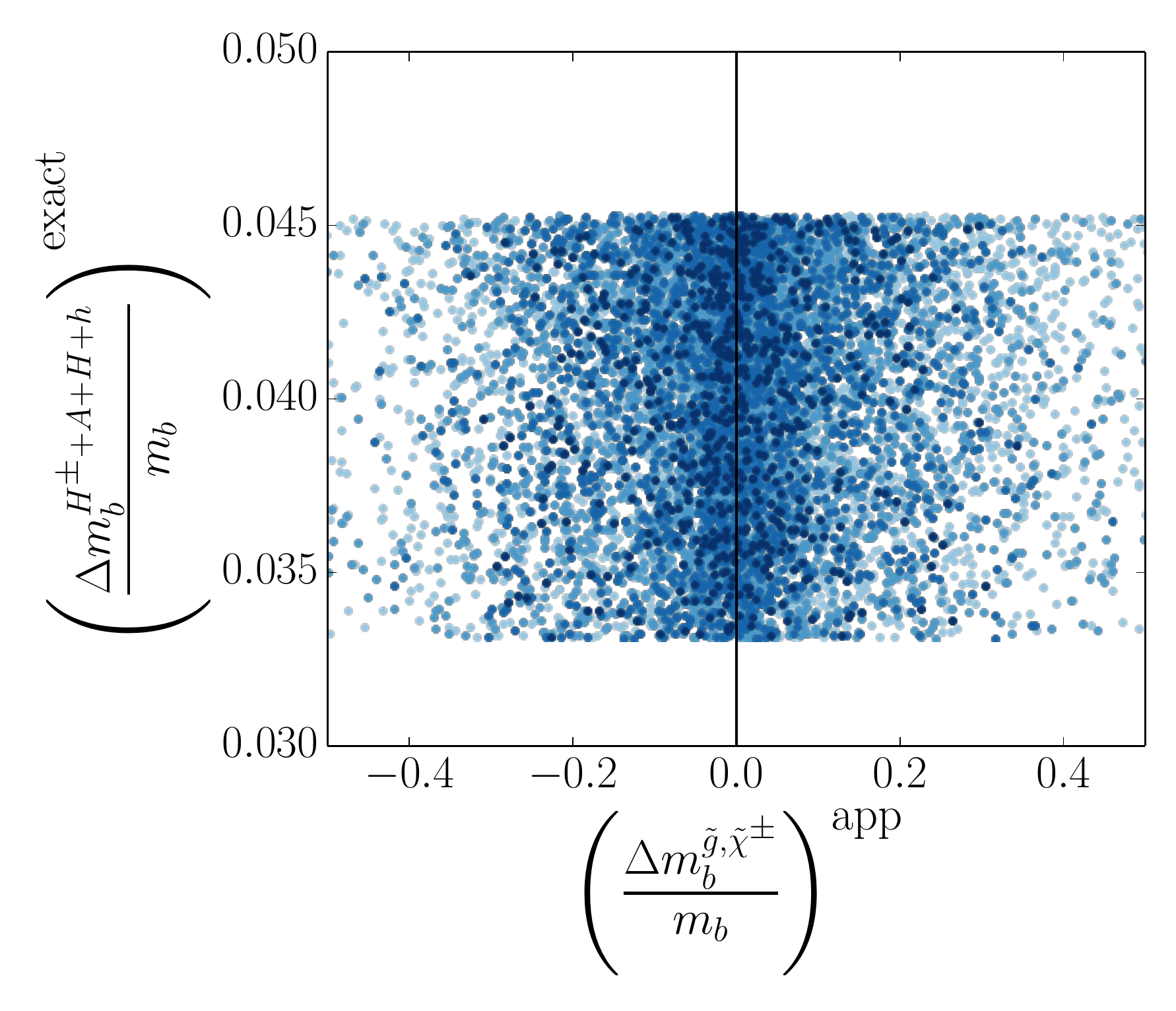} }}%
	\qquad
	\subfloat[Neutralinos]{{\includegraphics[width=0.46\textwidth, clip]{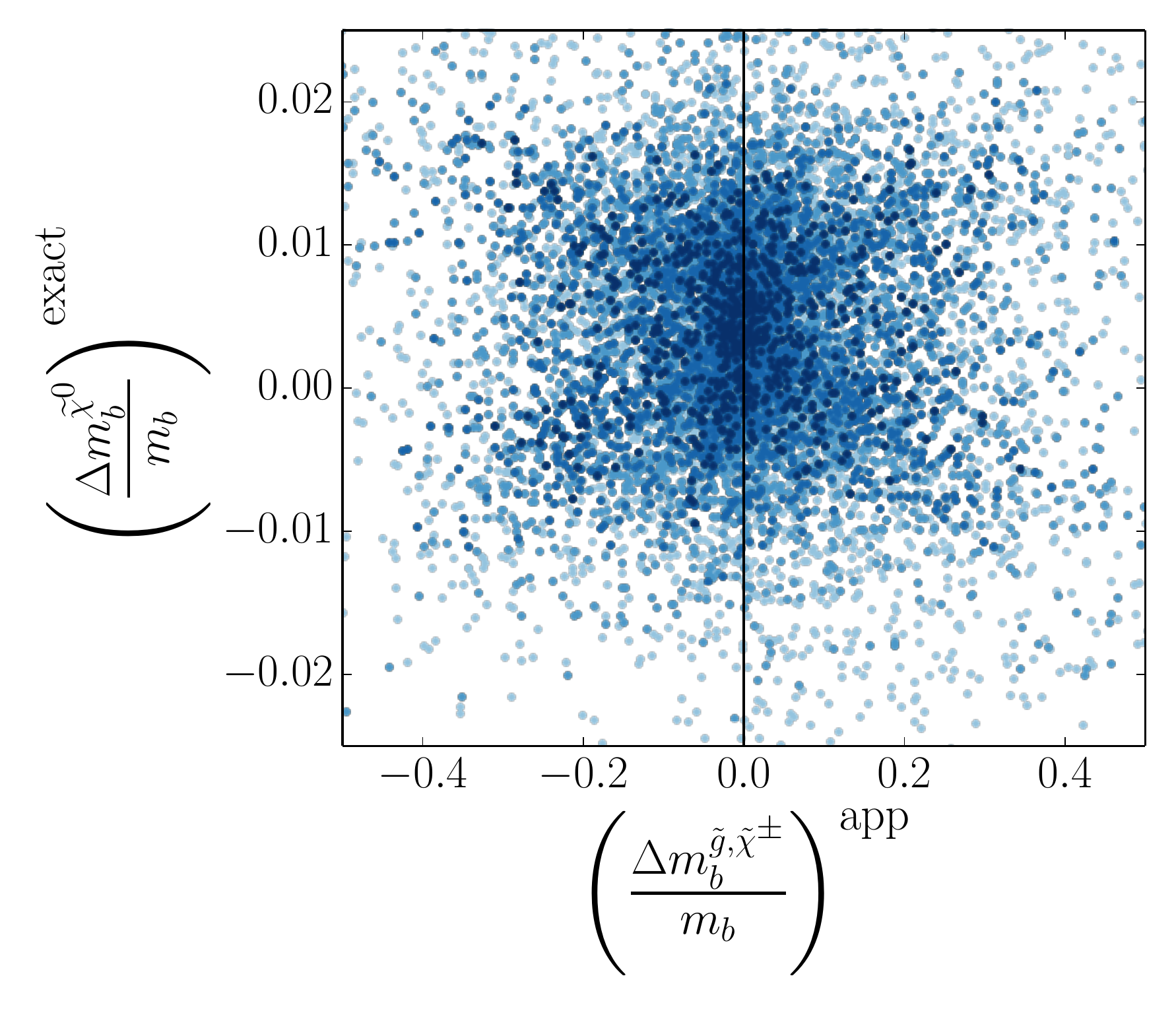} }}%
	\subfloat[W,Z]{{\includegraphics[width=0.46\textwidth, clip]{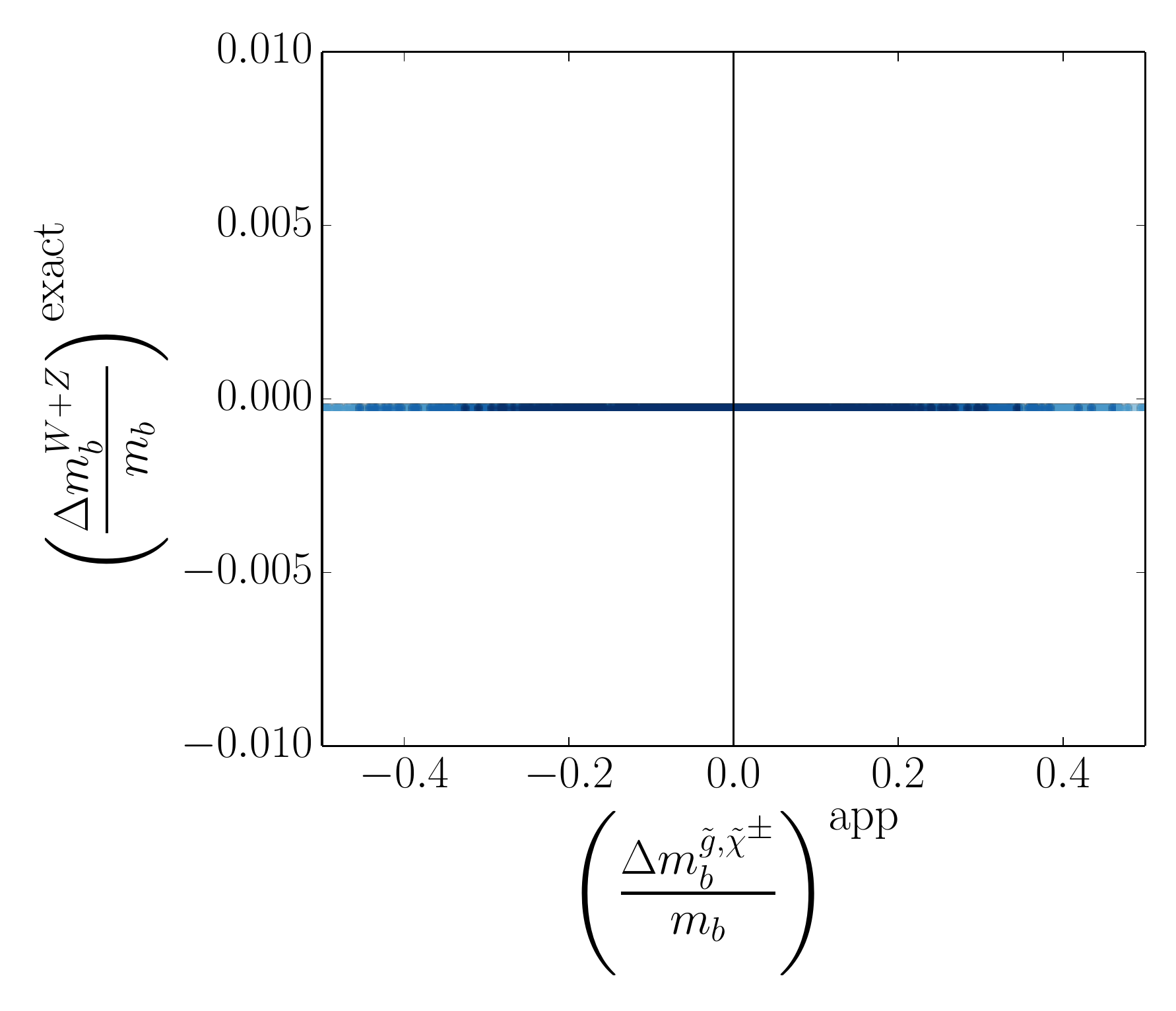} }}%	
	\caption{\small The plots show the relative size of the total approximate correction in comparison with the corrections from the contributions of the (top) Higgses, (bottom left) neutralinos, and (bottom right) $W$, and $Z$.}%
	\label{fig:others}%
	\end{figure}

	\ref{fig:others} shows the size of these quantities relative to the total approximate correction. We find that in the heavy squark regime the neutralino contribution is typically $\le$1\%. Furthermore, the $W$ and $Z$ contributions are very close to 0 for all points. This leaves the contributions from the Higgses, which give a correction of $\sim$4\% for all points. Thus, in the heavy squark regime in which the correction to the bottom quark mass given by~\ref{Eq:common-app} is small, the contribution from the Higgses should not be ignored. Note that the contributions from the Higgses are not $\tanb$-enhanced contributions~\cite{Pierce:1996zz}. The implications of this statement will be discussed in~\ref{Sec:consequences}.
	
% % % % % % % % % % % % % % % % % % % % % % % % % % % % % % % % % % % % % % % % %
% % % % % % % % % % % % % % % % % % % % % % % % % % % % % % % % % % % % % % % % %

	\section{Consequences \label{Sec:consequences}}
	In the previous section, we compared the magnitude of the SUSY threshold corrections to the bottom quark mass. Particularly, we have shown that the various approximations made to obtain the common form in~\ref{Eq:common-app} all seem to be valid approximations with the exception of neglecting the $B_1^{\widetilde{g}}$ terms in the gluino-sbottom contribution and possibly the contributions from the Higgses. In this section, we will highlight some of the consequences of including these terms in the corrections to the bottom quark mass.
	
	\subsection*{Fits to the bottom quark mass}
	A good choice of scale to integrate out the massive SUSY particles is the $M_Z$ scale. At the $M_Z$ threshold one then has to match the value of $m_b$ before and after integrating out the massive states. This leads to the relation
	\be
	m_b(M_Z)^{\rm SM} = m_b(M_Z)^{\rm MSSM}\left(1 + \Delta m_b/m_b\right)
	\label{eq:thresholds}
	\;.\ee
	$m_b(M_Z)^{\rm below}$ can be determined by taking the value of $m_b(m_b) = 4.19$ GeV and running it to the $M_Z$ scale. This is evaluated using the \code{RunDec} package to be $m_b(M_Z)^{\rm below} = 2.69$ GeV. The hope then is that the right choice of bottom Yukawa coupling and the appropriate set of SUSY boundary conditions at some UV scale will give rise to the necessary $m_b(M_Z)^{\rm above}$ and $\Delta m_b/m_b$ to satisfy~\ref{eq:thresholds}.
	
	When fitting the bottom quark mass, it is common to use the full, exact one-loop correction. This is done in most numerical spectrum calculators, such as \code{SOFTSUSY}~\cite{Allanach:2014nba} and \code{SPheno}~\cite{Porod:2003um}. Physical interpretations are often based however on the approximate formula given in~\ref{Eq:common-app}. As was shown in the previous section, additional terms, namely the $B_1^{\widetilde{g}}$ terms from the gluino-sbottom contributions and the contributions from the Higgses, should also be included for a full description. These ``missing" terms contribute $\sim$12\% to the correction. In~\ref{Sec:intro} the conditions for obtaining an appropriate SUSY threshold correction to the bottom quark mass in models with third family Yukawa unification were determined by an interpretation of the approximate formula given in~\ref{Eq:common-app}. We revisit this scenario here to offer a more accurate interpretation. 
	
	In models with third family Yukawa unification, the SUSY threshold corrections to the bottom quark typically need to be $-\mathcal{O}$(few)\%. For $\mu>0$, the common interpretation is $A_t$ needs to be large and negative in order for the chargino-stop contribution to overcome the $B_0^{\widetilde{g}}$ term from the gluino-sbottom contribution. By including the ``missing" terms, which are positive,	we see that the size of $A_t$ is underestimated when the approximate form of the corrections is used to interpret the size of the parameters. This is particularly true when the squarks are heavy. In this regime, the chargino-stop term is suppressed but the ``missing" terms are not and so $A_t$ must be quite large to overcome both the suppression by the heavy stops and also the positive contribution from the missing terms. It has been pointed out in earlier works that light Higgsinos are disfavoured in Yukawa unified GUTs~\cite{Baer:2012jp, Anandakrishnan:2013tqa}, and this can be traced back to~\ref{fig:ch-muM2}, where we see that the corrections from the chargino are small for small $\mu$ and do not compensate for the large gluino corrections. 
	
	For $\mu<0$, the common interpretation is that the parameters need not be large since the terms in~\ref{Eq:common-app} already have the needed minus sign. Including the ``missing" terms introduces a positive contribution (these terms are not proportional to $\mu$) that is relatively large and so $A_t$ and/or $M_{\widetilde{g}}$ must be larger than expected in order to overcome the additional contributions.

% % % % % % % % % % % % % % % % % % % % % % % % % % % % % % % % % % % % % % % % %
% % % % % % % % % % % % % % % % % % % % % % % % % % % % % % % % % % % % % % % % %	

    \subsection*{Higgs couplings to the bottom quark}
   The MSSM predicts four new physical Higgs states in addition to the light CP-even (SM like) Higgs boson. The coupling of the Higgs bosons to the bottom quark depends on the MSSM parameters, particularly, $\tanb$. In addition, the couplings also depend on the bottom quark threshold corrections and the effect of these corrections have been the subject of many works especially in the large $\tanb\ $ regime~\cite{Carena:1999py, Guasch:2001wv, Belyaev:2002eq, Guasch:2003cv, Crivellin:2009ar, Crivellin:2010er, Crivellin:2012zz, Baer:2011af, Carena:2012rw, Dittmaier:2014sva, Carena:2013qia}. The low energy effective Lagrangian coupling the bottom quark with the up- and the down-type Higgs bosons in the MSSM including the supersymmetric threshold corrections can be written as
   \be
   {\mathcal L}_{\rm eff} = -\lambda_b^0 \bar{b}_R^0 \left[(1 + \Delta_1) \phi^0_d + \Delta_2  \phi_u^{0*} \right] b_L^0 +\text{h.c.}
   \;,\ee
	
   where
   \bea
   \phi_d^0 &=& \frac{1}{\sqrt{2}} \left( v_d + H \cos \a - h \sin \a + i A \sin \b -i G^0 \cos \b \right) \\
   \phi_u^0 &=& \frac{1}{\sqrt{2}} \left( v_u + H \sin \a + h \cos \a + i A \cos \b +i G^0 \sin \b \right)
   \;.\label{Eq:bare-eff-lag}
   \eea    
    
   Here $\Delta_2$ represents the coupling of the bottom quark to the ``wrong" Higgs, which is generated by the radiative effects discussed in this paper. The corrections 
   to the coupling of the bottom quark to the down-type Higgs are represented by $\Delta_1$. The $\Delta_2$ interactions are $\tanb$-enhanced while the $\Delta_1$ corrections are not. The expression in~\ref{Eq:bare-eff-lag} must be matched to the renormalized Lagrangian given by~\cite{Guasch:2003cv}
   \be
   {\mathcal L}_{\rm eff} = -\lambda_b \bar{b}_R \left[\phi^0_d + \frac{\Delta_b}{\tanb}  \phi_u^{0*} \right] b_L +\text{h.c.}
   \;,\ee   
   yielding the relations
   \bea
   \lam_b = \lam_b^0(1+\Delta_1) \\ \nn\\
   \frac{\Delta_b}{\tanb} = \frac{\Delta_2}{1+\Delta_1}
   \;.\label{Eq:deltas}
   \eea
   
\begin{figure}[!]
   \thisfloatpagestyle{empty}
   \centering
   \includegraphics[width=0.54\textwidth, clip]{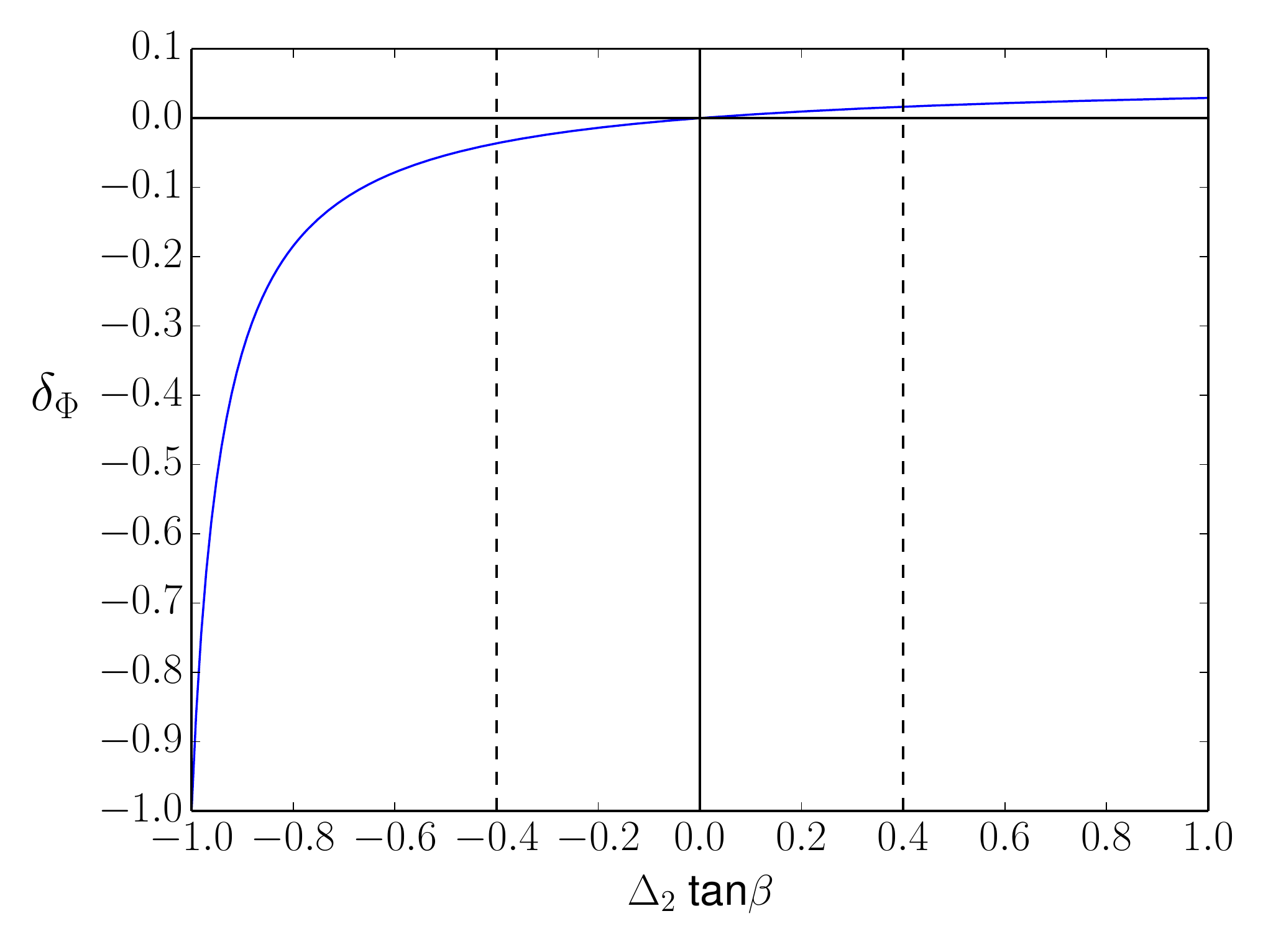}
   \caption{\small The plot shows the relative size of $\delta_2$, the correction to the heavy Higgs-bottom coupling ignoring the $\Delta_1$ contribution, and $\delta_{12}$, the correction to this coupling including the $\Delta_1$ contribution. The parameter $\delta_\Phi$ is defined in the text. The region within the vertical dashed lines is where most of the points from the parameter scan lie.}
   \label{fig:deltas}
   \end{figure}

   Consider the gluino contribution in the approximate form of the threshold corrections given by~\ref{Eq:common-app}. The $\mu$-term, which is proportional to $\tanb$, is included in $\Delta_2$ while the $A_b$-term is included in $\Delta_1$. The $\Delta_1$ correction is typically found to be $\mathcal{O}$(1)\% and is therefore often neglected~\cite{Guasch:2003cv}. We point out here that neither the $B_1^{\widetilde{g}}$ terms from the gluino-sbottom contribution nor the contributions from the Higgses are proportional to $\tanb$, and therefore they enhance $\Delta_1$ by $\sim$12\%. By considering the forms of the effective couplings of the Higgses to the bottom quark, we can determine if this enhancement translates to a nonnegligible correction. The effective couplings are given by~\cite{Carena:1999py}
   \bea
   \tilde{g}_b^h &=& \frac{g^h_b}{(1 + \Delta_b)} \left(1 - \frac{\Delta_b}{\tana \tanb} \right)\\
   \tilde{g}_b^H &=& \frac{g^H_b}{(1 + \Delta_b)} \left(1 + \frac{\Delta_b}{\cota \tanb} \right)\\
   \tilde{g}_b^A &=& \frac{g^A_b}{ (1 + \Delta_b)} \left(1 - \frac{\Delta_b}{\tan^2 \beta} \right)
   \;,\eea
   where $g^{h,H,A}_b$ are the tree level couplings. In the decoupling limit, $\tana \ra -\cotb$ and we obtain
   \bea
   \tilde{g}_b^h &=& g^h_b \\
   \tilde{g}_b^H &=& \frac{g^H_b}{ (1 + \Delta_b)} \left(1 - \frac{\Delta_b}{\tan^2 \beta} \right) \simeq \frac{g^H_b}{ (1 + \Delta_b)}\\
   \tilde{g}_b^A &=& \frac{g^A_b}{ (1 + \Delta_b)} \left(1 - \frac{\Delta_b}{\tan^2 \beta} \right) \simeq \frac{g^A_b}{ (1 + \Delta_b)}
   \;.\eea
   We therefore only need to determine the extent to which $\Delta_1$ affects the size of the factor $(1+\Delta_b)^{-1}$. From~\ref{Eq:deltas}, the factor may be written as
   \be
   \frac{1}{ 1 + \Delta_b}=\frac{1+\Delta_1}{1+\Delta_1+\Delta_2\tanb} \equiv \delta_{12}
   \;.\ee
   Let us define $\delta_2\equiv (1+\Delta_2\tanb)^{-1}$ and $\delta_\Phi$ to be the relative change between ignoring $\Delta_1$ and including it,
   \be
   \delta_\Phi \equiv \frac{\delta_{12}-\delta_2}{\delta_2}
   \;.\ee
   By setting $\Delta_1=0.12$, $\delta_\Phi$ can be plotted as a function of $\Delta_2\tanb$ as shown in~\ref{fig:deltas}. For positive values of $\Delta_2\tanb$, the relative change is never more than 6\%. Unless $\Delta_2\tanb$ is $\mathcal{O}(1)$, the relative correction to the heavy Higgs couplings is only a few percent. The effect of including $\Delta_1$ can be more drastic if $\Delta_2\tanb$ is negative. As $\Delta_2\tanb$ approaches $-\mathcal{O}(1)$, the relative change increases quickly to the nearly the same magnitude. Such large, negative values of $\Delta_2\tanb$ may be a more extreme case however. For most values of $\Delta_2\tanb$ obtained in the parameter scan $(<40\%)$, the relative change is again only a few percent. Thus unless the magnitude of the $\tanb$-enhanced corrections to the bottom quark mass are $\mathcal{O}(1)$ it is safe to neglect the $\Delta_1$ correction to the couplings of the bottom quark with the heavy Higgses.\footnote{It is expected that the LHC and ILC will be able to measure Higgs couplings to within a few percent~\cite{Peskin:2013xra,Han:2013kya}. It will then be necessary to include the $\Delta_1$ corrections.} Note that in calculating the $B_1^{\widetilde{g}}$ contribution to $\Delta_1$ we take $Q=M_Z$. If the scale is chosen to be higher, then $B_1^{\widetilde{g}}$ would be smaller and the relative change, $\delta_\Phi$, would be more suppressed.
   
% % % % % % % % % % % % % % % % % % % % % % % % % % % % % % % % % % % % % % % % %
% % % % % % % % % % % % % % % % % % % % % % % % % % % % % % % % % % % % % % % % %

%	\newpage
	\section{Conclusions \label{Sec:conclusions}}
	
	We have examined the validity of common approximations of the SUSY threshold corrections to the bottom quark mass. To avoid model dependency, we chose to work in the context of the pMSSM and performed a parameter scan to survey a large region of parameter space. In particular we considered large $\tanb$ and squark masses of $\mathcal{O}(\text{few})$ TeV. This choice is motivated by the absence of any newly discovered colored particles at the LHC. 	

	Comparing the full, exact one-loop expression to the common approximate form, we found for each point that the full expression is larger than the approximate expression by $\sim$12.5\%. The main sources of the discrepancy were determined to be the  contributions from the wave function renormalization coming from the gluino-sbottom diagrams ($\sim$8\%) and the contributions from the Higgses ($\sim$4\%), both of which are often neglected.	
	
	The consequences of an invalid approximation for the bottom quark threshold corrections were discussed for fits to the bottom quark mass and for the effective Higgs couplings to the bottom quark. We found that using the common approximation to determine the size of SUSY parameters needed to obtain desired bottom quark threshold corrections leads to an underestimation of the parameters. As for the effective Higgs couplings, including the oft-neglected contributions leads to a modification of $\mathcal{O}$(few)\% for nearly all points from the parameter scan. Thus the common approximation for the bottom quark threshold correction remains quite accurate for low energy bottom-Higgs phenomenology, even in the heavy squark regime.

	{\bf Acknowledgements:} We thank Michael Spira and Jaume Guasch for clarifying how the threshold corrections contribute to the Higgs couplings to bottom quarks and pointing out earlier works. We also thank Borut Bajc, St\'{e}phane Lavignac, and Timon Mede for useful comments.	AA is supported by the Ohio State University Presidential Fellowship. SR is supported by DOE/ DE-SC0011726. 
	
% % % % % % % % % % % % % % % % % % % % % % % % % % % % % % % % % % % % % % % % % % %
% % % % % % % % % % % % % % % % % % % % % % % % % % % % % % % % % % % % % % % % % % %

	\newpage
    \begin{appendices}
    \section{Gluino-sbottom}

\begin{figure}[h]
	\centering
	\subfloat[\footnotesize  -i $B_{LR}^x$]{
	\includegraphics[clip, width=0.4\textwidth]{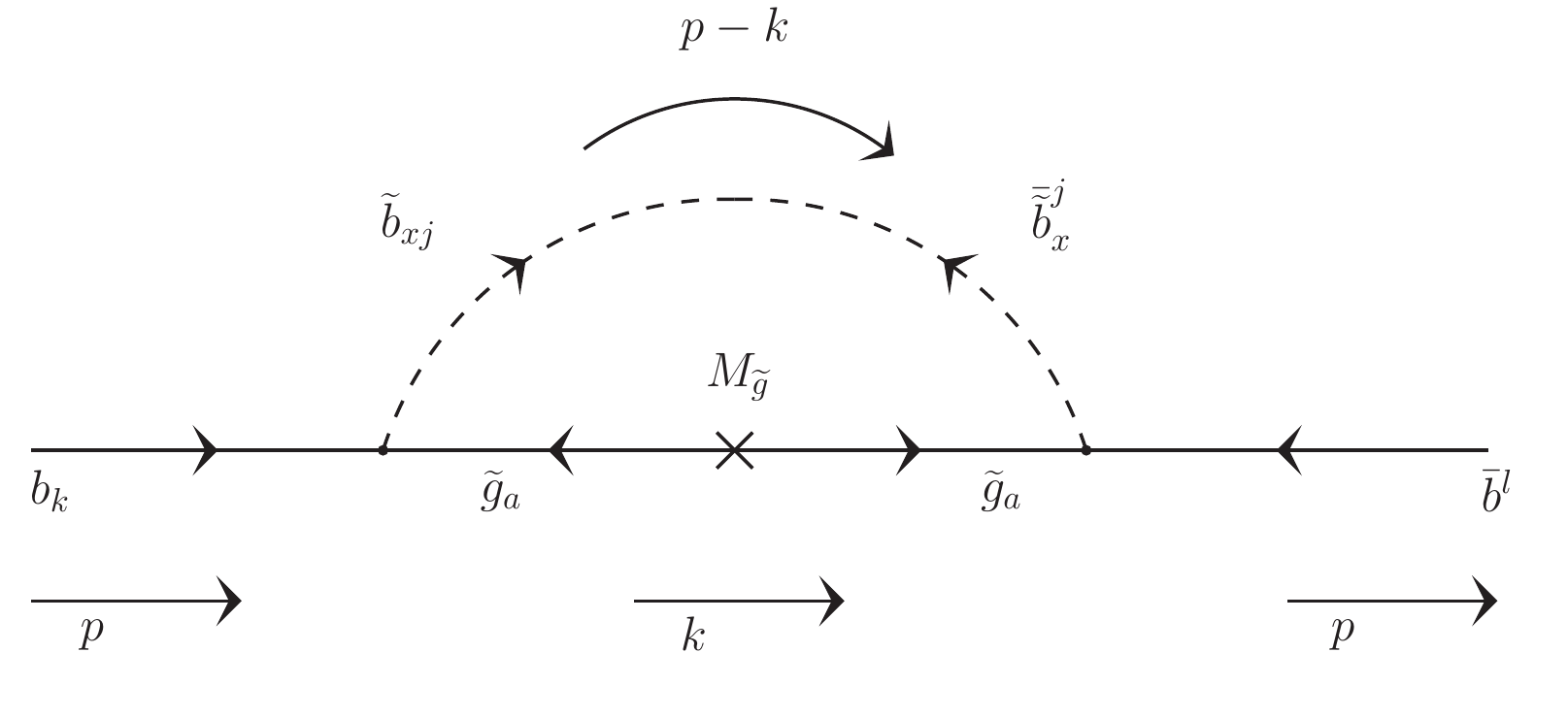}
	%\label{fig:FG-1}
	}\\
	\subfloat[\footnotesize -i $p \cdot \bar{\sigma} A_{L}^x$]{
	\includegraphics[clip,  width=0.4\textwidth]{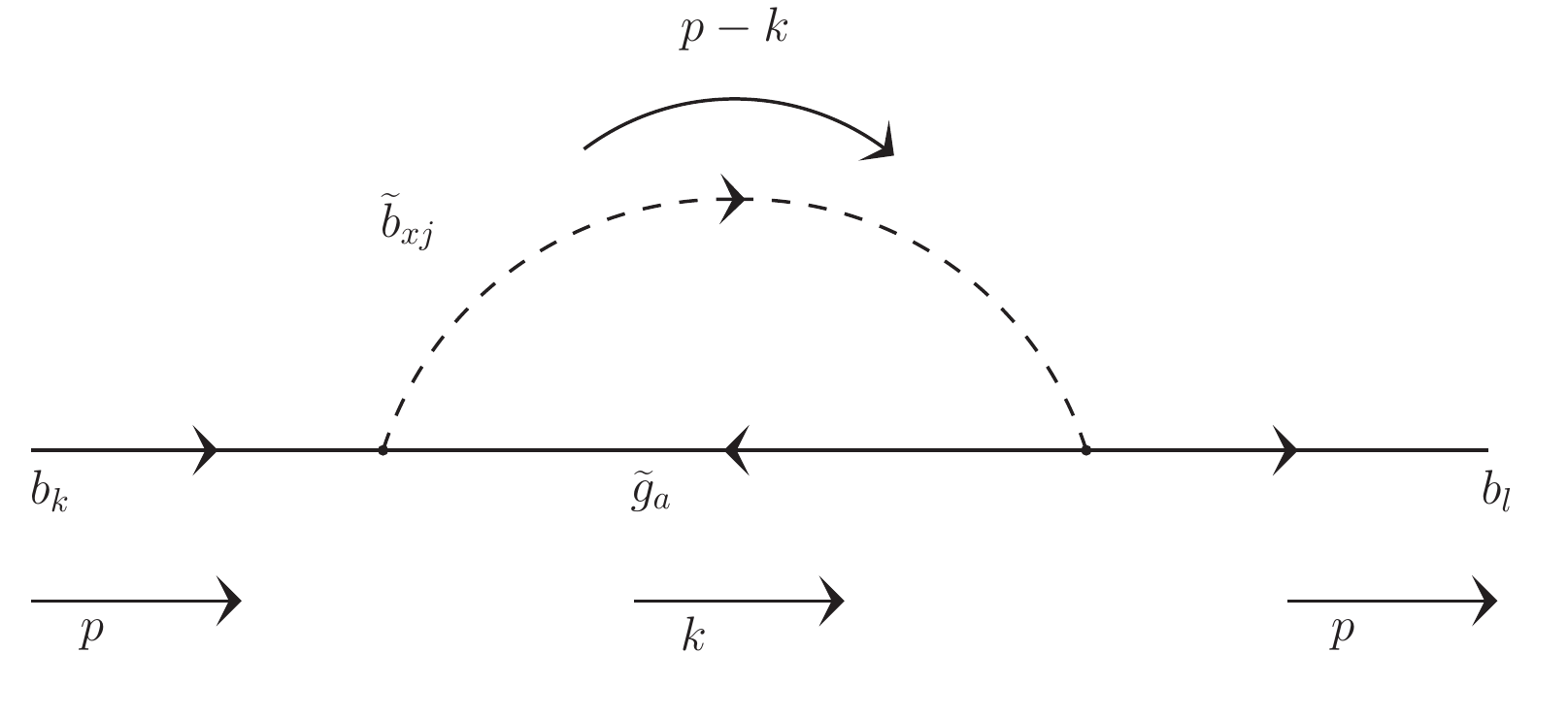}
	%\label{fig:FG-1}
	}
	\subfloat[\footnotesize -i $p \cdot \sigma A_{R}^x$]{
	\includegraphics[clip,  width=0.4\textwidth]{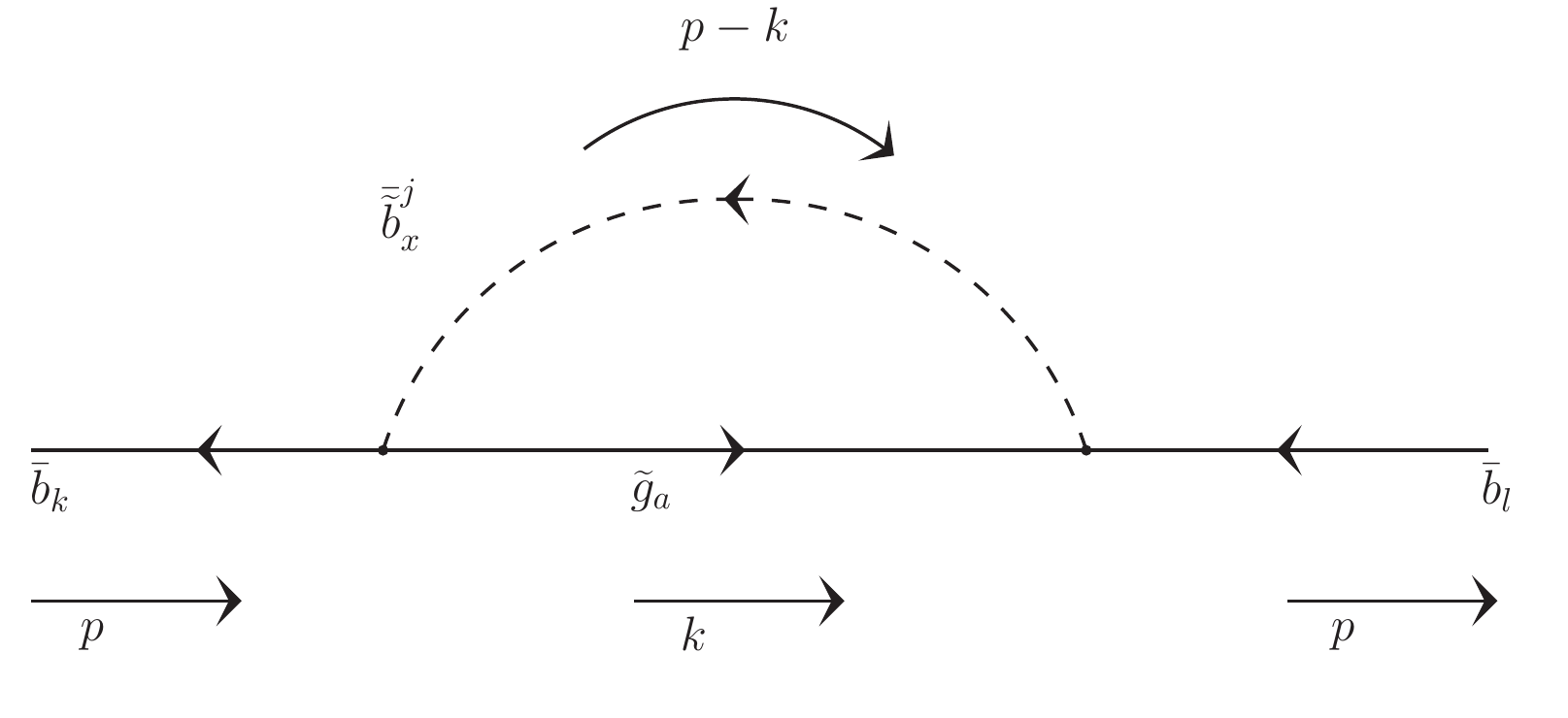}
	%\label{fig:FG-1}
	}
	\caption{\footnotesize Gluino-sbottom loops that give corrections to the inverse propagator of the bottom quark.}
	\label{gluinocorrections}
	\end{figure}

    Gluinos couple with the down-type squarks and quarks proportional to the SU(3) gauge coupling $g_3$ and hence contribute large corrections to the bottom quark mass. The corrections are dominant when the squarks belong to the third family since the inter-generational mixings between the squarks are typically (and by assumption in this study) small. We will now calculate the individual diagrams shown in~\ref{gluinocorrections} considering the contributions from the two bottom squarks.
    %There are three types of these diagrams that yield corrections to the bottom mass:
	
	The three diagrams correct the inverse propagator
	\be
	S(p) = \frac{i}{\slashed{p}-m-\Sigma(p)}
	\;,\ee
	
	where $-i \Sigma $ is the sum of the three diagrams in~\ref{gluinocorrections}:
	\be
	-i \Sigma (p) =  -i B_{LR} -i p \cdot \bar{\sigma} A_L -i p . \sigma A_R
	\;.\ee
	
	The Lagrangian after including the corrections from the diagrams can be written as
	\be
	\mathcal{L} = b^* i \slashed{D} b (1 - A_L) + \bar{b}^* i \slashed{D} \bar{b} (1-A_R) + \bar{b} b (m_{b0} + B_{LR})
	\;.\ee
	
    By rescaling $b$ and $\bar{b}$ by $\frac{1}{\sqrt{1-A_L}}$ and $\frac{1}{\sqrt{1-A_R}}$, respectively, the corrected bottom quark mass can be written as
	\bea
	m_b &=& \frac{m_{b0} + B_{LR}}{\sqrt{1-A_L}\sqrt{1-A_R}} \nonumber \\
	&\simeq& m_{b0} + B_{LR} + \frac{m_{b0}}{2} (A_L + A_R) \nonumber \\
	\Ra \Delta m_b = m_b - m_{b0} &=& B_{LR} + \frac{m_{b0}}{2} (A_L + A_R)
	\label{delmb}
	\;.\eea

We evaluate the loop integrals in each of the diagrams in~\ref{gluinocorrections}:
\bea
-i B_{LR}^x &=& \left( i \sqrt{2} g_3 \Gamma_R^x T_l^{aj} \right) \int \frac{d^4 k}{(2 \pi)^4} \left[\frac{i \mg}{k^2 - \mg^2 } \right]
\left( -i \sqrt{2} g_3 (\Gamma_L^x)^{\dagger} T_j^{ak} \right) \left[ \frac{i}{(p-k)^2 -  m_{\tilde{b}_x}^2 } \right] \nonumber \\
 &=& -\frac{8}{3} g_3^2 \Gamma_R^x \left(\Gamma_L^x \right)^{\dagger} \int \frac{d^4 k}{(2 \pi)^4} \frac{\mg}
{ \left( k^2 - \mg^2 \right) \left( (p-k)^2 -  m_{\tilde{b}_x}^2 \right) }  \nn \\[2pt]
-i p \cdot \bar{\sigma} A_{L}^x &=& \left(-i \sqrt{2} g_3 \Gamma_L^x T_l^{aj} \right) \int \frac{d^4 k}{(2 \pi)^4} \left[\frac{i k. \bar{\sigma}}
{k^2 - \mg^2 } \right]
\left( -i \sqrt{2} g_3 (\Gamma_L^x)^{\dagger} T_j^{ak} \right) \left[ \frac{i}{(p-k)^2 -  m_{\tilde{b}_x}^2 } \right] \nonumber \\
 &=& -i \frac{8}{3} g_3^2 \Gamma_L^x \left(\Gamma_L^x \right)^{\dagger} \int \frac{d^4 k}{(2 \pi)^4} \frac{i k \cdot \bar{\sigma}}
{\left( k^2 - \mg^2 \right) \left( (p-k)^2 -  m_{\tilde{b}_x}^2 \right) }  \nonumber \\[2pt]
-i p \cdot \sigma A_{R}^x &=& \left(i \sqrt{2} g_3 \Gamma_R^x T_l^{aj} \right) \int \frac{d^4 k}{(2 \pi)^4} \left[\frac{i k. \sigma}
{k^2 - \mg^2 } \right]
\left( i \sqrt{2} g_3 (\Gamma_R^x)^{\dagger} T_j^{ak} \right) \left[ \frac{i}{(p-k)^2 -  m_{\tilde{b}_x}^2 } \right] \nonumber \\
 &=& -i \frac{8}{3} g_3^2 \Gamma_R^x \left(\Gamma_R^x \right)^{\dagger} \int \frac{d^4 k}{(2 \pi)^4} \frac{i k \cdot \sigma}
{\left( k^2 - \mg^2 \right) \left( (p-k)^2 -  m_{\tilde{b}_x}^2 \right) }
\;.\eea

Using the standard definition of the Passarino-Veltman functions,
\bea
B_0(p, m_1, m_2) &=& 16 \pi^2 \int \frac{d^4 k}{i (2 \pi)^4} \frac{1}{(k^2 - m_1^2)((k-p)^2 - m_2^2)} \nn\\
p_\mu B_1(p, m_1, m_2) &=& 16 \pi^2 \int \frac{d^4 k}{i (2 \pi)^4} \frac{k_\mu}{(k^2 - m_1^2)((k-p)^2 - m_2^2)}
\label{passvelt}
\;,\eea
we get
\bea
B_{LR}^x &=& \frac{8}{3} \frac{g_3^2}{16 \pi^2} \Gamma_R^x \left(\Gamma_L^x \right)^{\dagger} \mg B_0 \left(p, \mg, m_{\tilde{b}_x}\right) \nonumber \\
A_{L}^x &=& -\frac{8}{3} \frac{g_3^2}{16 \pi^2} \Gamma_L^x \left(\Gamma_L^x \right)^{\dagger} B_1 \left(p, \mg, m_{\tilde{b}_x}\right) \nonumber \\
A_{R}^x &=& -\frac{8}{3} \frac{g_3^2}{16 \pi^2} \Gamma_R^x \left(\Gamma_R^x \right)^{\dagger} B_1 \left(p, \mg, m_{\tilde{b}_x}\right)
\;.\eea

Now we are ready to calculate the corrections to the bottom quark mass from the three diagrams as estimated in \ref{delmb}:

\bea
\Delta m_b^{\tilde{g}} &=& \sum_{x=1,2} B_{LR}^x + \frac{m_{b0}}{2} (A_L^x + A_R^x) \nonumber \\
&=& \frac{8}{3} \frac{g_3^2}{16 \pi^2} \sum_{x=1,2} \{ \Gamma_R^x \left(\Gamma_L^x \right)^{\dagger} \mg B_0 \left(p, \mg, m_{\tilde{b}_x}\right)
-\frac{m_b}{2} B_1 \left(p, \mg, m_{\tilde{b}_x}\right)(\Gamma_L^x \left(\Gamma_L^x \right)^{\dagger} +
\Gamma_R^x \left(\Gamma_R^x \right)^{\dagger}) \}\;. \nonumber\\
\eea
This is the exact expression for the one-loop threshold corrections to the bottom quark mass coming from the gluino-sbottom loops. In a full three family model, the $\Gamma_{L,R}$ are the $6 \times 3$ squark mixing matrices, and all the down-type squarks give rise to corrections to the bottom mass. Ignoring the off-diagonal elements that introduce the inter-generational mixing, we can consider a $2 \times 2$ block that mixes the two bottom squarks. The sbottom mixing matrix can be written as
\bea
\Gamma = \left( \begin{array}{cc}
\Gamma_L^1 & \Gamma_R^1  \\
\Gamma_L^2 & \Gamma_R^2 \end{array} \right) =
\left( \begin{array}{cc}
\cos \theta_b & \sin \theta_b  \\
-\sin \theta_b & \cos \theta_b \end{array} \right)
\;,\eea
such that
\bea
\left( \begin{array}{c}
\tilde{b}_1  \\
\tilde{b}_2 \end{array} \right) = \Gamma \left( \begin{array}{c}
\tilde{b}_L  \\
\tilde{b}_R \end{array} \right)
\;.\eea
Then, $\Delta m_b^{\tilde{g}}$ simplifies to
\bea
\Delta m_b^{\tilde{g}} = \frac{8}{3} \frac{g_3^2}{16 \pi^2} \left[ \frac{\sin 2\theta_b \mg}{2} \left( B_0 \left(p, \mg, m_{\tilde{b}_1}\right) - B_0 \left(p, \mg, m_{\tilde{b}_2}\right)  \right)  \right. \nn \\  -\left. \frac{m_b}{2} \left(B_1 \left(p, \mg, m_{\tilde{b}_1}\right) +
B_1 \left(p, \mg, m_{\tilde{b}_2} \right) \right)  \right]
\;.\eea

	\section{Chargino-stop}
\begin{figure}[h]
	\centering
	\subfloat[\footnotesize  -i $B_{LR}^x$]{
	\includegraphics[clip, width=0.4\textwidth]{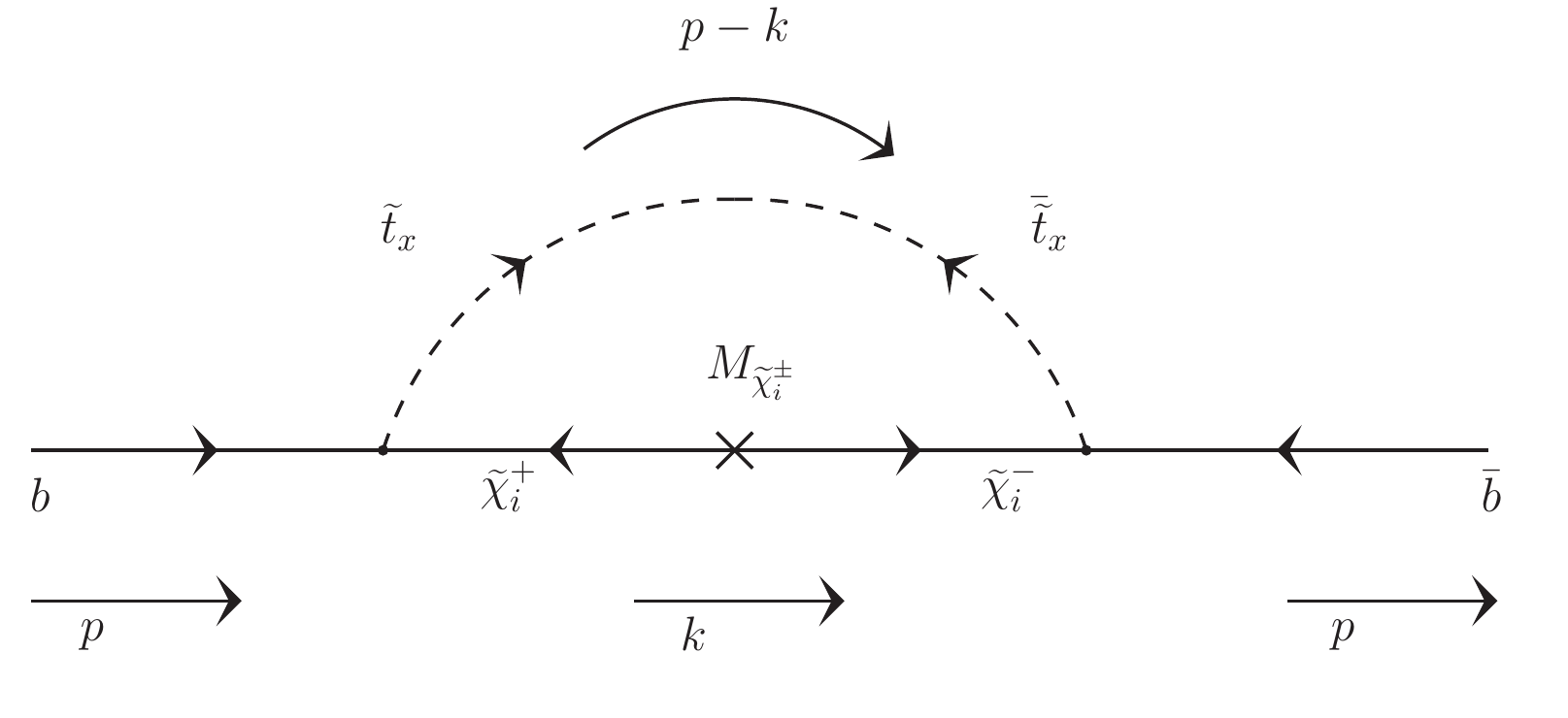}
	%\label{fig:FG-1}
	}\\
	\subfloat[\footnotesize -i $p \cdot \bar{\sigma} A_{L}^x$]{
	\includegraphics[clip, width=0.4\textwidth]{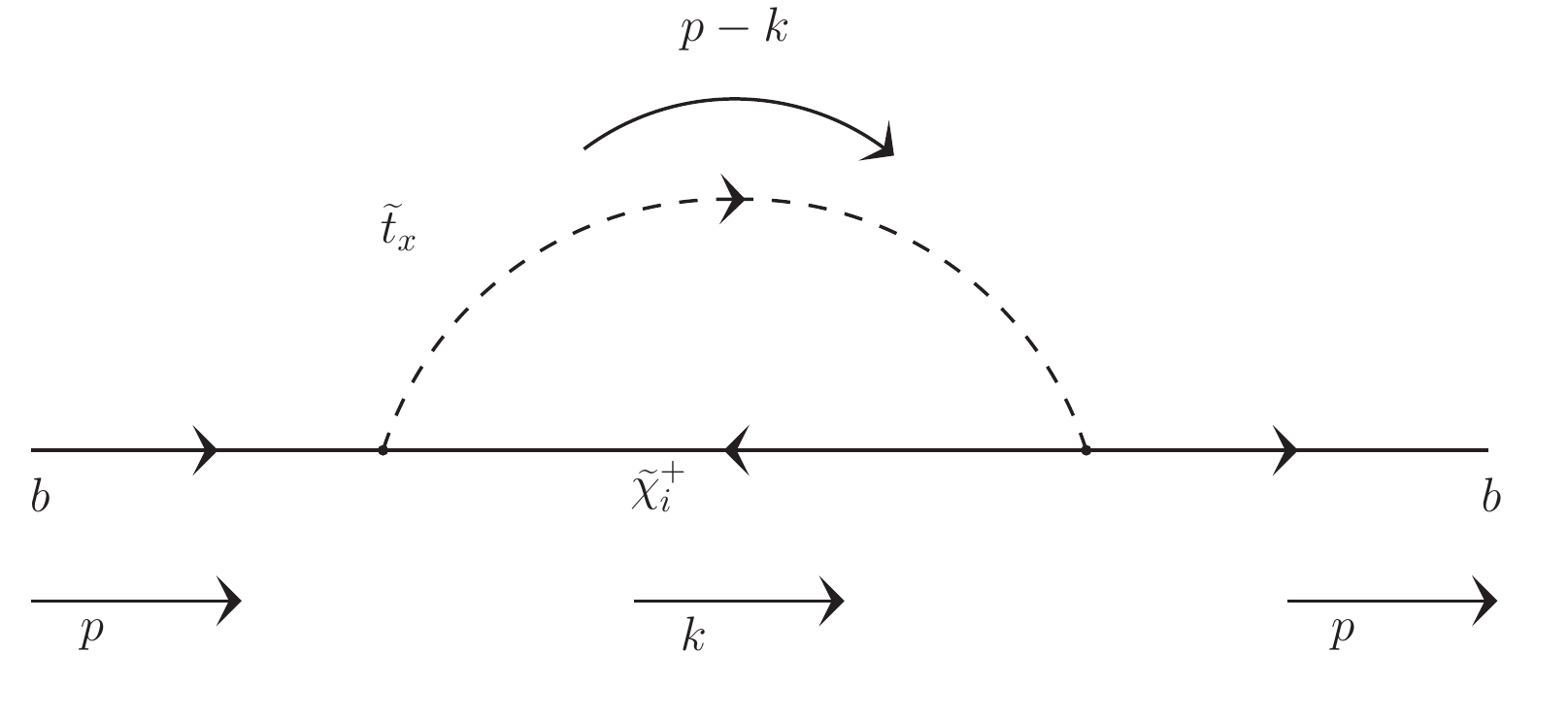}
	%\label{fig:FG-1}
	}
	\subfloat[\footnotesize -i $p \cdot \sigma A_{R}^x$]{
	\includegraphics[clip, width=0.4\textwidth]{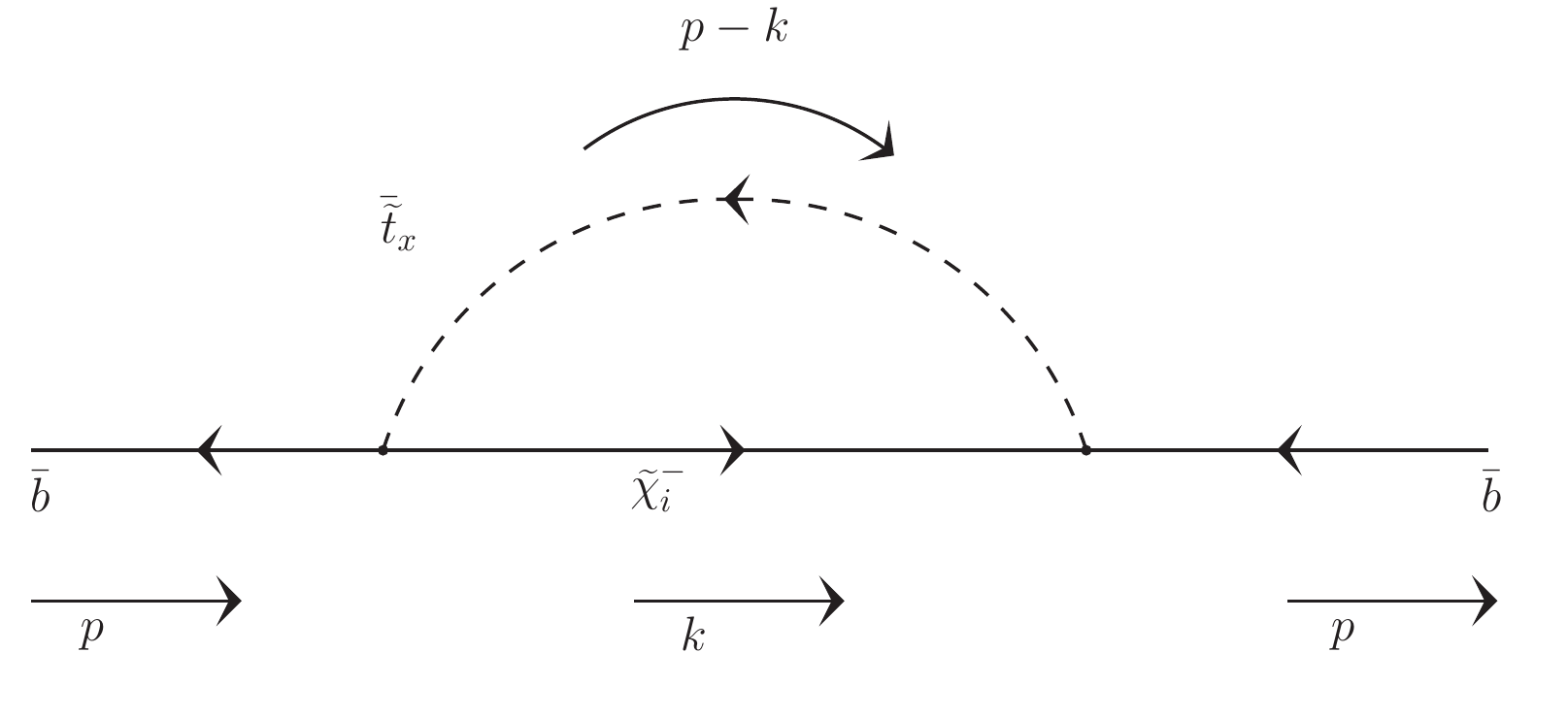}
	%\label{fig:FG-1}
	}
	\caption{\footnotesize Chargino-stop loops that give corrections to the inverse propagator of the bottom quark.}
	\label{charginocorrections}
	\end{figure}

	The charginos couple to the up-type squarks and down-type quarks proportional to the SU(2) coupling $g_2$ and the Yukawa couplings $\lam_{t,b}$ with strength depending upon their respective wino-higgsino composition. The corrections dominate when the squarks are from the third family due to CKM suppression of the contributions from the first two families of squarks. We calculate here the individual diagrams shown in~\ref{charginocorrections} considering the contributions from the two stop squarks. The calculation of the chargino-stop diagrams is similar to the calculation of the gluino-sbottom diagrams and yields
	%There are three types of these diagrams that yield corrections to the bottom mass:
	
\bea
-i B_{LR_i}^x &=&  \int \frac{d^4 k}{(2 \pi)^4} \left[i \bar{\Phi}^x_i \right] \frac{i \Mchipm{i}}{k^2 - (\Mchipm{i})^2 }  \left[i \Phi^x_i \right] \left[ \frac{i}{(p-k)^2 -  m_{\tilde{t}_x}^2 } \right]  \nn \\
 -i p \cdot \bar{\sigma} A_{L_i}^x &=& \int \frac{d^4 k}{(2 \pi)^4} \left[i \left(\Phi^x_i\right)^\dagger \right] \frac{i k \cdot \bar{\sigma}}{k^2 - (\Mchipm{i})^2 }
 \left[i \Phi^x_i \right] \left[ \frac{i}{(p-k)^2 -  m_{\tilde{t}_x}^2 } \right]  \nonumber \\
 -i p \cdot \sigma A_{R_i}^x &=& \int \frac{d^4 k}{(2 \pi)^4} \left[i \left(\bar{\Phi}^x_i\right)^\dagger \right] \left[\frac{i k \cdot \sigma}{k^2 - (\Mchipm{i})^2 } \right]
 \left[i \bar{\Phi}^x_i \right] \left[ \frac{i}{(p-k)^2 -  m_{\tilde{t}_x}^2 } \right]  \nonumber \\
 \Phi^x_i &=& \frac{\lambda_t}{\sqrt{2}} V^\dagger_{i2} \left(\Gamma^x_R\right)^\dagger - g_2 V^\dagger_{i1} \left(\Gamma^x_L\right)^\dagger \nn \\
 \bar{\Phi}^x_i &=& \frac{\lambda_b}{\sqrt{2}} U^\dagger_{i2} \Gamma^x_L
\;,\eea
where $\Phi$ and $\bar{\Phi}$ are the effective couplings of the bottom quark to a chargino mass eigenstate and a top squark. The gaugino fraction of the chargino couples proportional to the SU(2) gauge coupling $g_2$ and does not couple to the right-handed squarks. The Higgsino fraction of the charginos couples proportional to the Yukawa coupling of the top quark, $\lambda_t$. Once again, using the standard definition of the Passarino-Veltman function defined in~\ref{passvelt}, we get,
\bea
B_{LR_i}^x &=& -\frac{\bar{\Phi}^x_i \Phi^x_i \Mchipm{i}}{16 \pi^2} B_0(p,\Mchipm{i},m_{\tilde{t}_x}) \nn \\
A_{L_i}^x &=&  -\frac{ \left(\Phi^x_i\right)^\dagger \Phi^x_i}{16 \pi^2} B_1(p, \Mchipm{i}, m_{\tilde{t}_x}) \nn \\
A_{R_i}^x &=&  -\frac{ \left(\bar{\Phi}^x_i\right)^\dagger \bar{\Phi}^x_i}{16 \pi^2} B_1(p,\Mchipm{i},m_{\tilde{t}_x})
\;.\eea

The corrections to the bottom quark mass from the three diagrams in~\ref{charginocorrections} are then
\bea
\Delta m_b^{\tilde{\chi}^\pm_i} &=& \sum_{i=1}^2 \sum_{x=1}^2 B_{LR_i}^x + \frac{m_{b0}}{2} (A_{L_i}^x + A_{R_i}^x)
\label{fullchargino}
\;,\eea
where the sum runs over the two chargino mass eigenstates and the two stop eigenstates.

 \end{appendices}

% % % % % % % % % % % % % % % % % % % % % % % % % % % % % % % % % % % % %
%\newpage
\bibliographystyle{utphys} % use your favorite BIBTeX style
\bibliography{bibliography}

\end{document}